\documentclass[12pt]{article}
\usepackage{jheppub}

\usepackage{amsmath,bbm,array,amsfonts,graphicx,wrapfig,lscape,float,slashbox,multirow,longtable,rotating,subfigure,epstopdf}

\newcommand{\be}{\begin{equation}}
\newcommand{\ee}{\end{equation}}
\newcommand{\beq}{\begin{equation}}
\newcommand{\beql}[1]{\begin{equation}\label{#1}}
\newcommand{\eeq}{\end{equation}}
\newcommand{\ba}{\begin{array}}
\newcommand{\ea}{\end{array}}
\newcommand{\bea}{\begin{eqnarray}}
\newcommand{\beal}[1]{\begin{eqnarray}\label{#1}}
\newcommand{\eea}{\end{eqnarray}}
\newcommand{\ben}{\begin{enumerate}}
\newcommand{\een}{\end{enumerate}}
\newcommand{\bean}{\begin{eqnarray*}}
\newcommand{\eean}{\end{eqnarray*}}
\newcommand{\eref}[1]{(\ref{#1})}
\newcommand{\sref}[1]{\S\ref{#1}}

\newcommand{\fref}[1]{Figure \ref{#1}}
\newcommand{\btab}[1]{\begin{tabular}{#1}}
\newcommand{\etab}{\end{tabular}}

\newcommand{\comment}[1]{}

\newcommand{\qed}{\nobreak \ifvmode \relax \else
      \ifdim\lastskip<1.5em \hskip-\lastskip
      \hskip1.5em plus0em minus0.5em \fi \nobreak
      \vrule height0.75em width0.5em depth0.25em\fi}

\def\beqa{\begin{eqnarray}}
\def\eeqa{\end{eqnarray}}

\newcolumntype{C}[1]{>{\centering\arraybackslash}m{#1}}

\newcommand{\IX}{{\bf X}}

\def\makeatletter{\catcode`\@=11}
\makeatletter
\def\mathbox#1{\hbox{$\m@th#1$}}%
\def\math@ccstyles#1#2#3#4#5#6#7{{\leavevmode
     \setbox0\mathbox{#6#7}%
     \setbox2\mathbox{#4#5}%
     \dimen@ #3%
     \baselineskip\z@\lineskiplimit#1\lineskip\z@
     \vbox{\ialign{##\crcr
            \hfil \kern #2\box2 \hfil\crcr
            \noalign{\kern\dimen@}%
            \hfil\box0\hfil\crcr}}}}
\def\mathaccstyles{\math@ccstyles\maxdimen}
\def\maththroughstyles{\math@ccstyles{-\maxdimen}}
\def\unity%
{\maththroughstyles{.45\ht0}\z@\displaystyle {\mathchar"006C}\displaystyle 1}

\newcommand{\drawsquare}[2]{\hbox{%
\rule{#2pt}{#1pt}\hskip-#2pt
\rule{#1pt}{#2pt}\hskip-#1pt
\rule[#1pt]{#1pt}{#2pt}}\rule[#1pt]{#2pt}{#2pt}\hskip-#2pt
\rule{#2pt}{#1pt}}

\newcommand{\fund}{~\raisebox{-.5pt}{\drawsquare{6.5}{0.4}}~}
\newcommand{\antifund}{~\overline{\raisebox{-.5pt}{\drawsquare{6.5}{0.4}}}~}

\title{Bipartite Field Theories from D-Branes}

\author[a]{Sebasti\'an Franco}
\author[b]{and Angel Uranga}

\affiliation[a]{Institute for Particle Physics Phenomenology, Department of Physics\\
Durham University, Durham DH1 3LE, United Kingdom}
\affiliation[b]{Instituto de F\'isica Te\'orica IFT-UAM/CSIC \\
C/ Nicol\'as Cabrera 13-15, Universidad Aut\'onoma de Madrid, 28049 Madrid, Spain
}

\emailAdd{sebastian.franco@durham.ac.uk,angel.uranga@uam.es}

\abstract{We develop tools for determining the gauge theory resulting from a configuration of Type IIB D3-branes probing a non-compact, toric Calabi-Yau 3-fold, in the presence of additional flavor D7-branes with general embeddings. Two main ingredients of our approach are dimer models and mirror symmetry. D7-branes with general embeddings are obtained by recombination of elementary D7-brane constituents. These tools are then used to engineer a large set of Bipartite Field Theories, a class of 4d, $\mathcal{N}=1$ quantum field theories defined by bipartite graphs on bordered Riemann surfaces. Several explicit examples, including infinite families of models, associated to both planar and non-planar graphs are presented.}

\preprint{
\begin{flushright}IPPP/13/48\end{flushright} \vspace{-0.9cm}
\begin{flushright}DCPT/13/96\end{flushright} \vspace{-0.9cm}
\begin{flushright}IFT-UAM/CSIC-13-075\end{flushright} 
}

\begin{document}

\maketitle


\section{Introduction}

The string theory embedding of gauge theories often illuminates them from new perspectives, leading for instance to the geometrization of dualities and of other non-trivial dynamical properties. One possible scenario involves realizing quantum field theories on branes probing singular, non-compact geometries. This approach provides a bridge leading to gravity dual descriptions \cite{Maldacena:1997re,Gubser:1998bc,Witten:1998qj},  which allow the study of gauge theories at strong coupling in terms of supergravity. In addition, branes at singularities are the main ingredient of local approaches to string phenomenology \cite{Aldazabal:2000sa,Berenstein:2001nk,Verlinde:2005jr}. 

The case of Type IIB configurations of D-branes probing toric Calabi-Yau (CY) 3-folds is an example in which the map between geometry and quantum field theory can be controlled in exquisite detail. The key reason for this is the correspondence between these theories and dimer models \cite{Hanany:2005ve,Franco:2005rj,Franco:2005sm}, also known as brane tilings.

In this paper we extend the understanding of D-branes over toric CY 3-folds by developing a comprehensive framework for the inclusion of flavor D7-branes, i.e. D7-branes wrapping non-compact 4-cycles inside the CY. Part of our work is based on ideas originally introduced in \cite{Franco:2006es} and \cite{Forcella:2008au}, which we extend in various directions.

Recently, Bipartite Field Theories (BFTs), a new class of 4d $\mathcal{N}=1$ gauge theories, defined by a bipartite graphs on (bordered) Riemann surfaces, were introduced in \cite{Franco:2012mm}.\footnote{In \cite{Xie:2012mr}, a closely related class of theories was defined, which coincides with the ones in \cite{Franco:2012mm} in the absence of  boundaries. The reader is referred to \cite{Franco:2013pg} for a discussion of the connection between the two types of theories.}  These theories include and generalize brane tilings. Remarkably, some of these theories also have deep connections to integrable systems \cite{GK,Franco:2011sz,Eager:2011dp,Franco:2012hv,Amariti:2012dd} and on-shell diagrams in $\mathcal{N}=4$  super Yang-Mills \cite{ArkaniHamed:2012nw,Franco:2012mm,Franco:2012wv,Amariti:2013ija}.  

In the second part of this article, we exploit our tools for studying general configurations of D-branes over toric CY singularities to engineer a large class of BFTs. An alternative approach for embedding similar theories in string theory has been presented in \cite{Heckman:2012jh}.

The organization of this paper is as follows. In \sref{section_background} we review background material, including dimer models, their mirror interpretation and tadpole cancellation for D-branes at singularities. We also initiate the discussion of basic configurations of flavor D7-branes. In \sref{section_BFTs} we briefly review the definition of BFTs. In \sref{sec:d7s-full}, we introduce tools for determining the gauge theories arising from configurations involving flavor D7-branes with general embeddings. In \sref{section_BFTs_from_branes}, we describe the application of these ideas to engineer large classes of BFTs.  We conclude in \sref{section_conclusions}.

\bigskip

\section{Background}

\label{section_background}

In this section we review the connection between D3-branes on toric CY 3-folds and dimer models, the mirror of such configurations, the inclusion of flavor D7-branes with simple embeddings and the constraints following from tadpole cancellation. 

\bigskip

\subsection{D-Branes on Toric Calabi-Yau 3-Folds and Dimer Models}

\label{section_dimer_models}

The quiver gauge theories on D3-branes over toric CY 3-folds are described by dimer models, also denoted brane tilings \cite{Hanany:2005ve,Franco:2005rj,Franco:2005sm}. A brane tiling is a bipartite graph on a 2-torus. A bipartite graph is a graph in which nodes can be colored black or white, such that edges only join nodes of different color. The translation between the dimer and the gauge theory is given by:

\begin{table}[htt!!]
\begin{center}
\begin{tabular}{|l|l|}
\hline
{\bf Dimer} & {\bf Quiver} \\ \hline \hline
Face $i$ & Gauge group $U(N_i)$ \\ \hline
Edge $e_{ij}$ between faces & Chiral multiplet $X_{ij}$ in the bifundamental $(\fund_i,\antifund_j)$ \\ $i$ and $j$ & representation, oriented  clockwise around white nodes \\ &  and counterclockwise around black  nodes \\ 
\hline 
$k$-valent node joining edges  & Monomial  $X_{i_1i_2} X_{i_2i_3}\ldots X_{i_ki_1}$
in the superpotential \\  
$e_{i_1i_2} e_{i_2i_3}\ldots e_{i_ki_1}$  & involving $k$ chiral  multiplets, with sign ($+/-$) for \\ & (white/black) nodes \\ \hline
\end{tabular}
\caption{Dictionary connecting dimer models and the corresponding gauge theories.\label{dictionary}}
\end{center}
\end{table}

The ranks for the gauge groups associated to faces in the dimer can be different. Different rank assignments correspond to possible choices of fractional D3-branes and are constrained by tadpole cancellation as explained in \sref{sec:anomaly}. For a detailed discussion of dimer models, we refer the reader to \cite{Franco:2005rj}.

\bigskip

\subsection{The Mirror}

\label{section_D3s_mirror}

One of the main tools we will use for analyzing D-branes on a toric CY 3-fold is the mirror configuration, which was first discussed in connection with dimer models in \cite{Feng:2005gw}. The mirror geometry is a $\Sigma_w \times \mathbb{C}^*$ fibration over the $w$ complex plane, given by
\begin{eqnarray}
P(x,y) & = & w  \nonumber \\
u \, v & = & w \, .
\end{eqnarray}
There is a Riemann surface $\Sigma_w$ corresponding to $P(x,y) = w$, for every point $w$. Here $P(x,y)=\sum a_{n_1,n_2} x^{n_1} y^{n_2}$ is the characteristic polynomial of the geometry under consideration, i.e. there is a term in $P(x,y)$ for every point in the toric diagram with position $(n_1,n_2)$. 

\begin{figure}[h]
 \centering
 \begin{tabular}[c]{ccc}
 \epsfig{file=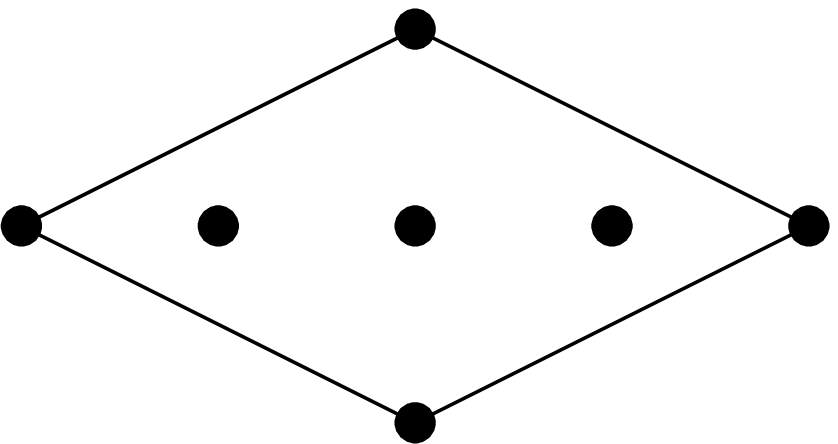,width=0.3\linewidth,clip=} & \ \ \ \ \ &
\epsfig{file=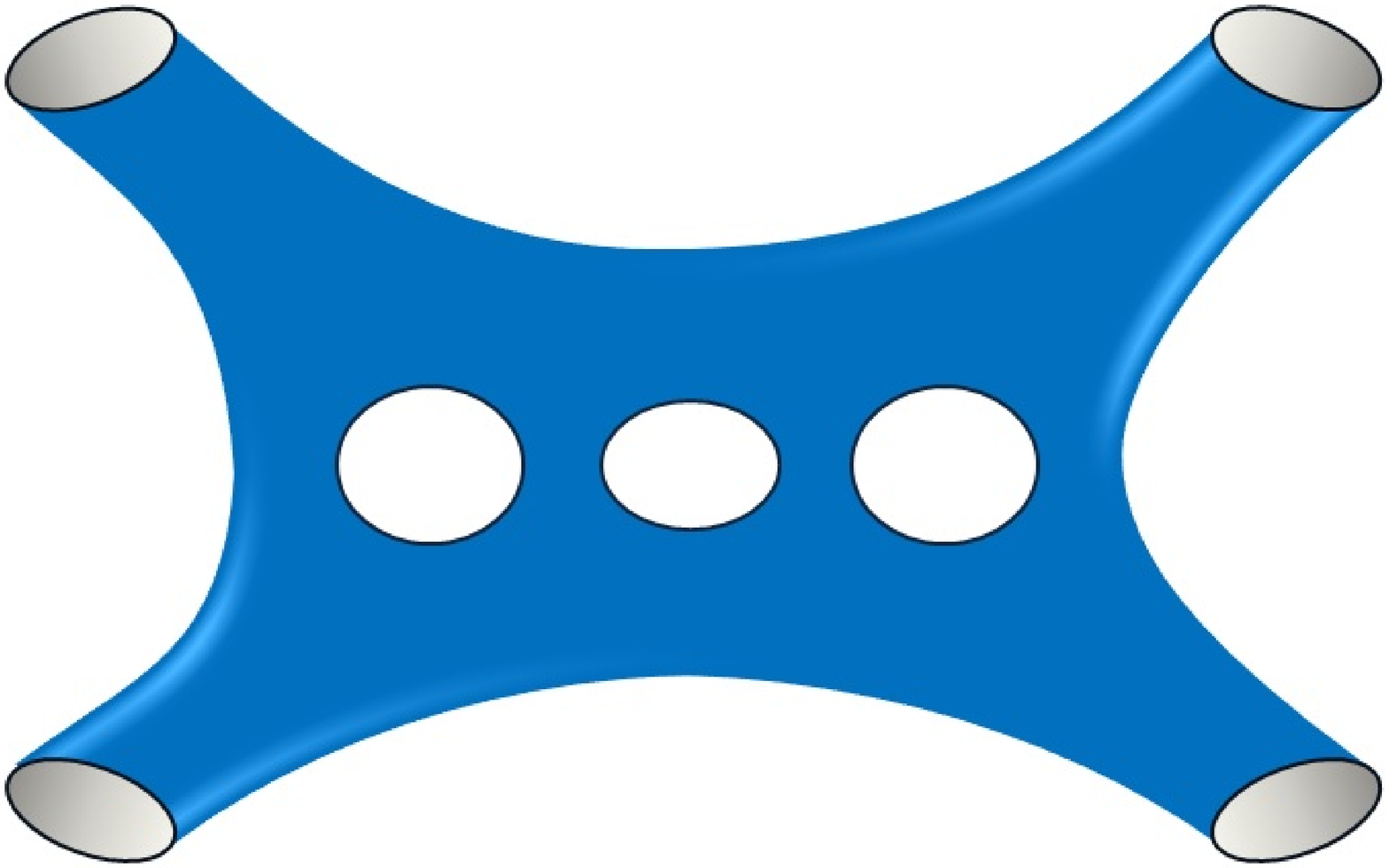,width=0.27\linewidth,clip=} \\ 
(a) & & (b)
 \end{tabular}
\caption{{\textit  a) Example of a toric diagram for a CY singularity. b) The corresponding Riemann surface $\Sigma$, defined by $P(x,y)=0$, in the mirror geometry.}}
\label{toric_diagram_vs_sigma} 
\end{figure}

The main aspects concerning the physics D-branes in this geometry are captured by the Riemann surface $\Sigma$ sitting at the origin $w=0$, i.e. the surface defined by $P(x,y)=0$. The skeleton of $\Sigma$ is the $(p,q)$ web associated to the CY \cite{Aharony:1997ju,Aharony:1997bh}, which in turn is the graph dual to its toric diagram. The genus and number of punctures of $\Sigma$ are hence equal to the number of internal points and perimeter of the toric diagram, respectively. \fref{toric_diagram_vs_sigma} shows an example.

\begin{figure}[h]
\begin{center}
\includegraphics[width=9.5cm]{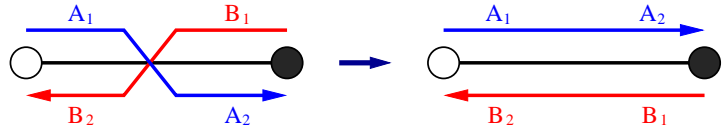}
\caption{\textit{The untwisting map.} Its action on zig-zag paths, here represented in double line notation.}
\label{untwisting}
\end{center}
\end{figure}

The dimer model allows us to understand the connection between the original CY and its mirror as follows. We define zig-zag paths as connected oriented sequences of consecutive edges which turn maximally right at black nodes and maximally left at white nodes; they can be usefully depicted as oriented paths crossing edges by the middle, as in the upcoming figures.  The mirror geometry is revealed upon using the {\it untwisting map}, which acts on zig-zag paths of the dimer as schematically shown in \fref{untwisting}. Namely, the zig-zag paths of the dimer model $G$ are organized as moving around faces in a new bipartite graph $\tilde{G}$.\footnote{In fact, $G$ and ${\tilde G}$ are the same graph, i.e. they contain the same edges and nodes. We give them different names in order to emphasize that they differ in their two dimensional embeddings into $T^2$ and $\Sigma$, respectively. These embeddings result in different sets of faces for each of them.} The latter turns out to define a tiling of the mirror Riemann surface $\Sigma$, with each face in $\tilde{G}$ corresponding to a puncture in $\Sigma$.

The untwisting map exchanges:

\medskip

\beq
\begin{array}{ccccc}
\mbox{{\bf $G$ on 2-torus}} & & & & \mbox{{\bf $\tilde{G}$ on $\Sigma$}} \\
\mbox{zig-zag path} & \ \ \ \ & \leftrightarrow & \ \ \ \ & \mbox{face (puncture)}\\
\mbox{face (gauge group)} & \ \ \ \ & \leftrightarrow & \ \ \ \ & \mbox{zig-zag path}
\end{array}
\nonumber
\eeq
\medskip
It is straightforward to see that a second application of the untwisting map to $\tilde{G}$ results in the original graph $G$. 

The physical meaning of the previous transformation is as follows. Every face in $G$ corresponds to a class of fractional D3-branes in the original CY. Fractional branes map to D6-branes wrapped over compact 3-cycles in the mirror. These 3-cycles project down to compact 1-cycles on $\Sigma$. More concretely, the D3-brane whose gauge groups is associated to a face of $G$, is mapped to the corresponding zig-zag path of ${\tilde G}$, c.f. the above table.

\begin{figure}[h]
\begin{center}
\includegraphics[width=12cm]{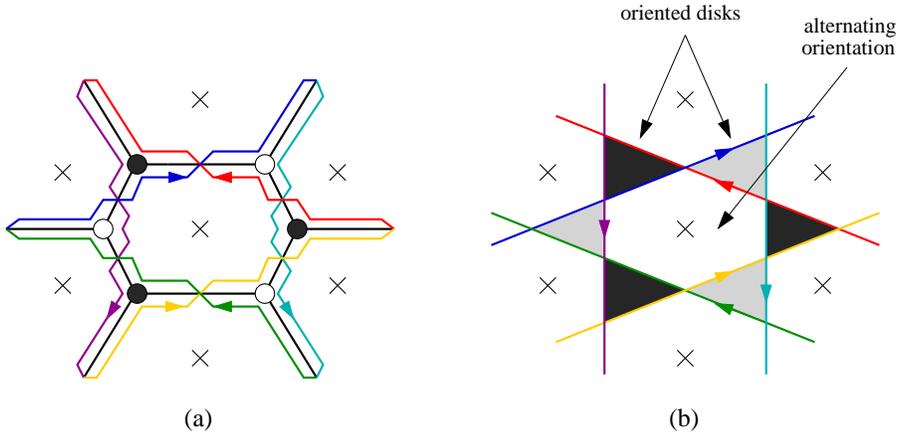}
\caption{a)	A piece of a bipartite graph $\tilde{G}$ in the mirror, with crosses indicating punctures. Zig-zag paths are shown in double line notation. b) The same configuration showing only the zig-zag paths, i.e. the D6-branes. Black and white nodes, which are shown here in grey for clarity, correspond to clockwise and counterclockwise oriented disks, respectively.}
\label{zig-zags_mirror}
\end{center}
\end{figure}

Since the gauge theory associated to a set of intersecting D6-branes depends just on its topology, it is possible to discuss the mirror just in terms of the zig-zag paths, as we often do in what follows. It is not necessary to refer to $\tilde{G}$, which can be reconstructed from them. \fref{zig-zags_mirror} gives an example illustrating the main properties of the connection between $\tilde{G}$ and its zig-zag paths:

\medskip

\begin{itemize}
\item {\bf Edge (chiral field):} intersection between two zig-zag paths, which supports open strings in the bifundamental of the corresponding gauge factors.
\item {\bf Node (superpotential term):} disk with an oriented boundary, which supports a worldsheet instanton mediating the interaction. The orientation is clockwise for white nodes and counterclockwise for black nodes.
\item {\bf Face (puncture):} disk whose boundary changes orientation at each intersection between zig-zag paths.
\end{itemize}

\medskip

\noindent Exactly the same observations apply for the original graph $G$ and its zig-zag paths. In this case, however, faces should be interpreted as gauge groups in the corresponding quiver gauge theory.

\begin{figure}[h]
\begin{center}
\includegraphics[width=7cm]{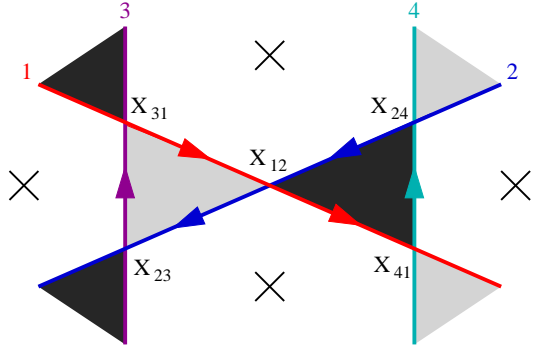}
\caption{Two adjacent oriented disks in the mirror, giving rise to a black and a white node connected by an edge. We indicate the bifundamental fields arising at the intersections between D6-branes. The white node corresponds to the $X_{12}X_{23}X_{31} $ term in the superpotential and the black one corresponds to $-X_{12}X_{24}X_{4}$.}
\label{zig-zags_relative_orientation}
\end{center}
\end{figure}

The order in which bifundamental fields are contracted to form gauge invariant superpotential couplings is determined by their orientation. In the original dimer $G$, this orientation is clockwise for white nodes and counterclockwise for black nodes, and it coincides with the orientation of the corresponding disks bounded by zig-zag paths. However in the mirror, as a result of untwisting, the orientation of gauge invariant contractions is clockwise for all nodes. This means that for black nodes this orientation is opposite to the one determined by zig-zag paths. \fref{zig-zags_relative_orientation} presents a simple example illustrating this behavior.

\bigskip

\subsection{Mirror Description of D7-Branes: Short Embeddings}

\label{section_D7s_short_embeddings}

Systems of D3-branes at singularities can be enriched by introducing additional D7-branes, spanning non-compact 4-cycles passing through the singular point. At the level of the field theory on the D3-brane, the D3-D7 open string sectors lead to the introduction of flavors for the diverse D3-brane gauge factors, in a pattern to be described later on. This has been exploited for the construction of Particle Physics models in \cite{Aldazabal:2000sa} (see e.g. \cite{Krippendorf:2010hj,Dolan:2011qu,Cicoli:2013zha} for other recent references), and in the context of flavored AdS/CFT (see e.g. \cite{Ouyang:2003df,Kuperstein:2004hy}, building on the original flavoring of ${\cal N}=4$ SYM in \cite{Karch:2002sh}).

Just like for D3-branes, the matter multiplets and superpotential couplings associated to D7-branes are more manifest in the mirror picture. 
In this section we describe some general aspects of the latter, in simple situations which we dub {\it short embeddings}. The detailed description of fairly more general configurations and their couplings is postponed to \sref{sec:d7s-full}.

\begin{figure}[h]
\begin{center}
\includegraphics[width=12cm]{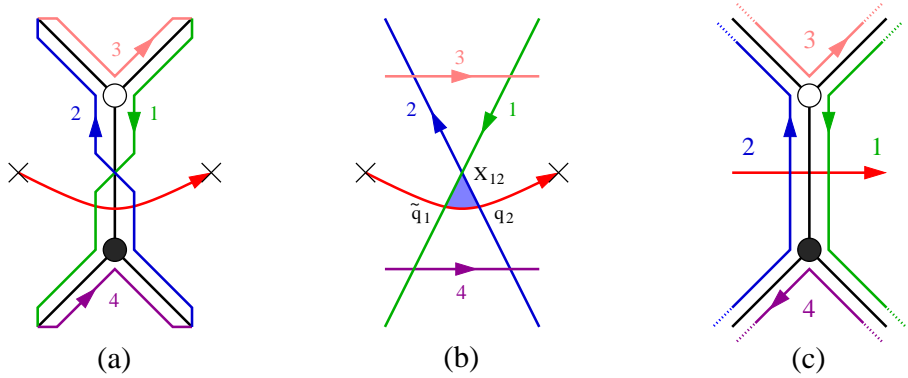}
\caption{\textit{Flavor D7-brane.} a) In the mirror, it is represented by an oriented 1-cycle connecting two punctures of $\Sigma$. b) The same configuration, omitting the underlying graph $\tilde{G}$. A pair of flavors, $\tilde{q}_1$ and $q_2$, arise at the intersection of the D7-brane with the corresponding zig-zag paths. The disk supporting the instanton that generates the superpotential $W_{3\, 7}= {\tilde q}_{1} X_{12} q_{2}$ is shown in blue. c) The corresponding piece of the original dimer, which is connected to the mirror by untwisting.}
\label{mirror_short_embedding}
\end{center}
\end{figure}

D7-branes on non-compact 4-cycles translate to D6-branes on non-compact 3-cycles in the mirror, which project onto non-compact 1-cycles on $\Sigma$, coming in and out through two different punctures. The simplest situation is that the 1-cycle crosses a single edge in the tiling ${\tilde G}$. It thus crosses the two corresponding zig-zag paths, namely intersects two D3-brane cycles, see \fref{mirror_short_embedding}. The intersections produce flavors, namely matter multiplets in the fundamental and antifundamental of the D3-brane gauge factors, due to the opposite orientation of the intersections. Incorporating the D7-brane groups as a global symmetry groups from the 4d field theory viewpoint, the multiplets transform in the corresponding bifundamental representations. We take the orientation of the D7-brane 1-cycle such that there is an oriented disk, which supports a worldsheet instanton generating a superpotential term with the schematic structure
\beqa
W_{3\,7}= {\tilde q}_{73} X_{33'} q_{3'7} .
\label{supo0}
\eeqa
The orientation of the cycle is opposite to that of the chiral multiplet for the corresponding edge in the dimer. In the original dimer, we will represent the D7-brane by an arrow across an edge, with its head and tail signaling the corresponding flavors, as shown in \fref{mirror_short_embedding}.c. We note that the D7-brane arrow is oriented opposite to the bifundamental to which the D3-D7 fields couple.

In principle, it is possible for a given pair of punctures to be connected by inequivalent paths, crossing different edges. This happens for D7-branes which wrap the same geometric 4-cycle but differ in their worldvolume gauge field. Specifically, the 4-cycle with the singular point removed is non-simply connected, and admits gauge fields with different holonomies. In orbifold examples, the choice of gauge holonomy at infinity in the 4-cycle (retraction of the 4-cycle minus the singular point) is described as the orbifold action on the Chan-Paton degrees of freedom. Regardless of the detailed description in the singular CY, the rules to read out the D3-D7 spectrum and interactions are as above, by simply choosing the appropriate field $X_{33'}$ in (\ref{supo0}).
These ideas were introduced and discussed in detail in \cite{Franco:2006es}, where they were illustrated using the $\mathbb{C}^3/\mathbb{Z}_3$ example, for which the D7-branes can be characterized using worldsheet CFT tools. 

An interesting connection should be noted at this point \cite{Forcella:2008au}. Consider a conical CY singularity $\IX_6$ with (real) base a 5d (Sasaki-Einstein) manifold ${\bf Y}_5$, as often done in the AdS/CFT context \cite{Klebanov:1998hh,Morrison:1998cs}. There is a direct relation between conical 4-cycles in $\IX_6$ and their (real) 3d bases in ${\bf Y}_5$. Under it, dibaryons of the D3-brane field theory, which are dual to D3-branes wrapped on supersymmetric 3-cycles in ${\bf Y}_5$, are naturally associated to holomorphic 4-cycles in $\IX_6$ (with `3-cycles' and `4-cycles' regarded in a generalized sense, as including data on the worldvolume gauge field background for the corresponding branes, as explained above) \cite{Gubser:1998fp,Berenstein:2002ke,Beasley:2002xv,Butti:2006au,Forcella:2007wk}. Thus, large sets of D7-branes wrapped on the holomorphic 4-cycles can be characterized in terms of properties of their associated dibaryon operator \cite{Forcella:2008au}. Jumping ahead a little bit, c.f. \sref{section_long_embeddings},  D7-branes can be associated to a general open path in the dimer, generated by concatenation of D3-D3 bifundamental fields ${\cal O}_{i_0i_n}=X_{i_0i_1}X_{i_1i_2}\ldots X_{i_{n-1}i_n}$, introducing flavors ${\tilde q}_{7 \, i_0}$, $q_{i_n 7}$, and with superpotential coupling ${\tilde q}_{7\, i_0} {\cal O}_{i_0i_n} q_{i_n 7}$.

The main purpose of this paper is to extend the set of D7-brane geometries which can be considered and used in applications. This requires a careful analysis of configurations including several different D7-branes, and of the properties of the corresponding D7-D7$'$ open string sectors. This leads to a rich set of possible D7-brane geometries, spectra and interactions, to be developed in \sref{sec:d7s-full}.

\bigskip

%
\subsection{Tadpole Cancellation}
\label{sec:anomaly}

In any string configuration including D-branes with a compact transverse space, an important microscopic consistency condition is cancellation of tadpoles for non-dynamical RR fields. Even in non-compact situations, cancellation of tadpoles must be imposed for RR potentials whose degree equals the dimensionality of the subspace on which they propagate. Considering the prototypical example of D3-branes at complex dimension-3 orbifold singularities, analyzed in \cite{Leigh:1998hj}, consistency requires RR tadpole cancellation in sectors twisted by orbifold elements having the origin as the only fixed point, since they produce RR 4-forms localized on the 4d subspace located at the singular point. On the other hand, one need not impose RR tadpole cancellation in sectors twisted by orbifold elements leaving fixed planes, since the corresponding twisted RR fields have 6d support, and their flux lines can escape to infinity. The resulting conditions constrain the allowed multiplicities of D-branes to be located in each face of the dimer diagram. In terms of the adjacency matrix $I_{ab}$, counting the net number of edges between faces $a$ and $b$ (counted with orientation) in the corresponding dimer, the conditions read
\beqa
\sum_b N_b I_{ab}=0 \quad {\rm for}\; {\rm all}\; a.
\label{rr-tadpole}
\eeqa
This can be expressed as the statement that the number of `fundamental' and `antifundamental' representations of the $a^{th}$ `gauge factor' are equal, with the peculiarity that it must be imposed even for faces which are empty ($N_a=0$), or have no non-abelian factor ($N_a=1$), or have no complex representations ($N_a=2$); hence the quotation marks. For the same reason, RR tadpole cancellation conditions imply, but may be slightly stronger than, the cancellation of non-abelian anomalies in the resulting 4d field theory.

The above discussion holds for general systems of D3-branes at singularities, in particular for dimer models, by simply taking the corresponding matrix $I_{ab}$. This can be shown by computation of the RR charges carried by the D-branes, by laboriously extrapolating the D-branes to large volume (which does not change topological charges) \cite{Wijnholt:2002qz,Herzog:2003zc,Hanany:2006nm}. Alternatively, and most conveniently for our purposes, it can be shown in the mirror picture, where the RR tadpole condition amounts to cancellation (in compactly supported homology) of the total homology class of the cycles wrapped by the mirror D6-branes (with multiplicity) \cite{Uranga:2002pg}. A further derivation for the above consistency condition follows from using brane probes (for instance, brane-antibrane pairs of the regular branes mentioned below) to test potential underlying inconsistencies \cite{Uranga:2000xp}.

The RR tadpole cancellation conditions also imply the cancellation of mixed $U(1)$ anomalies, by a 4d version of the Green-Schwarz mechanism (see \cite{Ibanez:1998qp} for orbifold theories); this moreover renders the $U(1)$ factors massive, so they decouple and disappear from the theory. It is worthwhile to emphasize that the cancellation mechanism works without imposing any further consistency conditions. Despite the disappearance of abelian factors, it is often useful to focus precisely on the case of abelian gauge theories with $N_a=1$, since many results of this simpler setup hold also in the more involved non-abelian case.

For dimer models, the RR tadpole cancellation conditions always admit the `regular D3-brane' solution of all ranks being equal, $N_a\equiv N$ for all $a$. This relates to the fact that such configuration is continuously connected (by moving onto the mesonic moduli space, i.e. moving the brane off the singularity) to the configuration of $N$ dynamical branes in the bulk of the CY, which is obviously consistent. In order to engineer BFTs with topologies other than the torus, e.g. the disk, we should consider configurations with empty faces in large regions of some underlying dimer. Although some such systems may be realized with only D3-branes, the class of modes which can be constructed is much richer if we also include D7-branes in the system, as advanced earlier. 

The RR tadpole cancellation conditions to systems of D3 and D7-branes at singularities, and their relation with (non-abelian and mixed) anomaly cancellation in the resulting 4d theory, are a generalization of the above discussion (see \cite{Aldazabal:1999nu,Aldazabal:2000sa} for orbifold examples). They are given by (\ref{rr-tadpole}), with the proviso that we let $b$ (but not $a$!) run also through the D7-brane stacks. These conditions are natural in the mirror picture, where both kinds of objects turn into D6-branes wrapping (either compact or non-compact) 3-cycles. The consistency conditions are the cancellation of compactly supported 3-homology charge \cite{Uranga:2002pg}. 

Therefore, for practical purposes in the remainder of the paper, the question of RR tadpole cancellation is addressed with the following rule:

\medskip
\begin{center}
\begin{tabular}{|c|}
\hline
The RR tadpole cancellation conditions can be taken as the `cancellation of \\ non-abelian anomalies' (i.e. equality of incoming and outgoing arrows, counted \\ with multiplicity), for all faces in the dimer, including those which are empty \\
\hline
\end{tabular}
\end{center}

\bigskip

\section{Bipartite Field Theories}

\label{section_BFTs}

Bipartite Field Theories (BFTs) are 4d, $\mathcal{N}=1$ gauge theories whose Lagrangian is defined by a bipartite graph on a Riemann surface, which might contain borders \cite{Franco:2012mm}. The translation between the graph and the field theory follows a natural generalization of the dictionary for dimer models given in Table \ref{dictionary}. Below we discuss the few new ingredients that can appear in these theories. We refer a reader to \cite{Franco:2012mm} for a thorough discussion of BFTs and \cite{Franco:2012mm,Franco:2012wv,Cremonesi:2013aba} for explicit examples beyond dimer models.

\begin{figure}[h]
\begin{center}
\includegraphics[width=12cm]{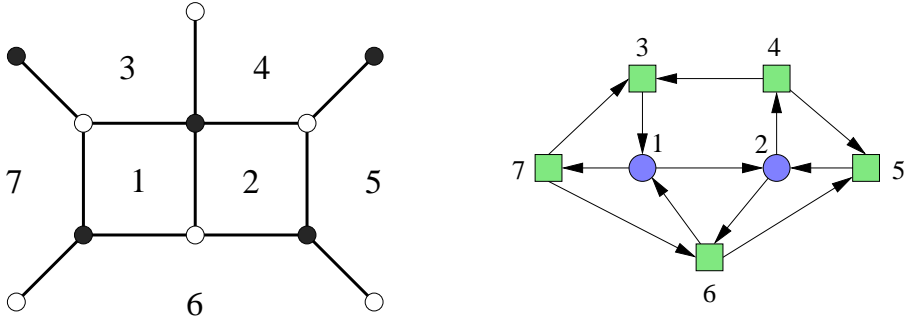}
\caption{\textit{Bipartite graph and its dual BFT.} Example of a BFT associated to a graph on a disk. Internal faces in the graph correspond to gauge symmetry groups, which are represented by blue circles in the quiver diagram. External faces correspond to global symmetries, which are represented by green squares.}
\label{BFT_example}
\end{center}
\end{figure}

The embedding into the Riemann surface is an important part in the specification of the BFT, since it determines a set of two-dimensional faces cut out by the graph. In the presence of boundaries, there are two possible types of faces: internal and external.  Both classes of faces correspond to $U(n_i)$ symmetry groups. However, only the ones associated to internal faces are gauged, while the ones for external faces correspond to global symmetries. 

Edges in the graph correspond to chiral multiplets in bifundamental or adjoint representations as in \sref{section_dimer_models}. Those associated to external legs are treated specially and are regarded as non-dynamical. In the string theory constructions of \sref{section_BFTs_from_branes}, this property will naturally follow from the fact that such fields have a higher dimensional support. 

Finally, external nodes, i.e. nodes sitting on boundaries are allowed to be connected to a single edge in graph and have no interpretation as a superpotential term.

As it will become clear in \sref{section_BFTs_from_branes}, the gauge theory on certain systems of D3/D7-branes at a CY singularity can be nicely encoded in the language of BFTs.

\bigskip

\section{General D7-Branes on Toric Calabi-Yaus}
\label{sec:d7s-full}

In this section we extend the set of D7-brane configurations which can be considered. The basic idea is to decompose general D7-branes into constituent ones with the type of simple embeddings discussed in \sref{section_D7s_short_embeddings} and recombine them by turning on non-zero vacuum expectation values (vevs) for bifundamental fields in the D7-D7' sectors. Hence, an important preliminary development is the analysis of configurations including several different D7-branes, of the corresponding D7-D7$'$ open string sectors, and their couplings to the D7-D3 and D3-D3 sectors. The determination of the corresponding rules can be carried out without caring about cancelation of tadpoles/anomalies, which can be dealt with in a subsequent stage. The explicit examples in \sref{section_BFTs_from_branes} illustrate the details of tadpole cancellation.

\bigskip

\subsection{D7-D7$'$ Sectors}

\label{section_77_sectors}

In this section we explain how to determine the spectrum between different D7-branes.

The main novelty in configurations including different D7-branes, is the possible appearance of new fields in the mixed D7-D7$'$ open string sectors. Since two non-compact 4-cycles generically intersect over a non-compact 2-cycle, these fields have 6d support and do not manifest as dynamical 4d fields. Still, their vevs can couple to the D3-brane gauge theory as external parameters, and so their determination is relevant to the 4d physics. 

Since the D7-D7$'$ fields are non-compact, their spectrum would seem not to be uniquely determined by the local geometry, since new modes can be brought from infinity. However, there is a non-trivial piece of the spectrum which is determined by the local geometry, and which can be regarded as arising from the compactly supported induced D-brane charges. This sector manifests in terms of D7-D7$'$ `spurion' chiral multiplets with non-trivial couplings to the 4d fields in the D3-D7 and D3-D3 sectors. Our techniques, which particularly exploit the mirror picture, allow an easy determination of this part of the D7-D7$'$ spectrum and its interactions. The resulting rules generalize results for flat space or orbifolds thereof, and are described in the following.

A D7-D7$'$ field extends between pairs of D7-branes with opposite orientations that `intersect' at a puncture in the mirror. The non-compactness of the puncture and the 1-cycles reflects the non-compact support of this sector. These fields can arise from the 6d intersections of D7-branes wrapped on different 4-cycles (two 1-cycles sharing a single puncture),  but also from the 8d volume of D7-branes wrapping the same 4-cycle but carrying different Chan-Paton gauge bundles (1-cycles sharing both punctures, but differing in their `interior'), c.f. \sref{section_D7s_short_embeddings} . Each such `puncture intersection' produces a D7-D7$'$ `spurion' 4d chiral multiplet, appearing in superpotential couplings to the 4d fields in D3-D7 and D3-D3 sectors. The models in \sref{section_BFTs_from_branes} are explicit examples of this rule, in which the D7-D7' spectrum can be independently determined using standard orbifold techniques.

The simplest configuration, which will often appear in the examples of \sref{section_BFTs_from_branes}, corresponds to two D7-branes $A$ and $B$ which give (opposite chirality) flavors to a common gauge group $i$ and share a puncture. This occurs whenever two D7-branes sit on two edges that are consecutive when going around the perimeter of a face in the original dimer. In this case, the D7-branes automatically share a zig-zag path in the dimer, i.e. a puncture in the mirror. A D7-D7$'$ field $Y_{AB}$ arises, with the following superpotential coupling to flavors 
\beq
W'_{\rm 3\, 7} = q_{iA}\, Y_{AB}\, \tilde{q}_{Bi}.
\label{W_73-33-37}
\eeq
The chirality of $Y_{AB}$ is such that this coupling is gauge invariant, i.e. it is set by the two flavors $q_{iA}$ and $\tilde{q}_{Bi}$ connecting the D7-branes to the common gauge group $i$.

\begin{figure}[h]
\begin{center}
\includegraphics[width=6cm]{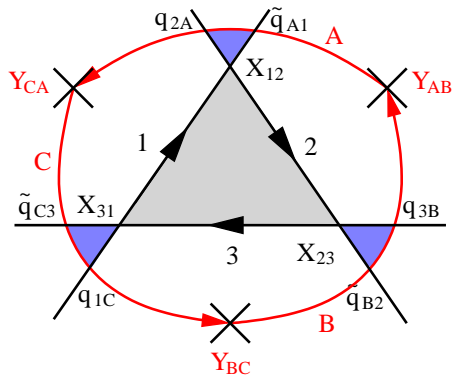}
\caption{D7-branes in the mirror for a set of punctures around a disk associated to a white node in the bipartite graph. Blue disks correspond to 73-33-37 couplings.}
\label{D7s_white_node}
\end{center}
\end{figure}

\begin{figure}[h]
\begin{center}
\includegraphics[width=14cm]{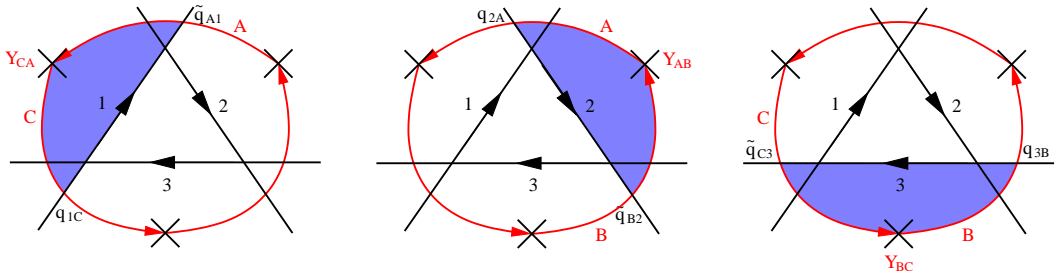}
\caption{Oriented disks giving rise to 37-77-73 superpotential couplings.}
\label{D7s_white_node_37-77-73}
\end{center}
\end{figure}

Let us consider the mirror in more detail. \fref{D7s_white_node} shows an oriented disk (shaded in grey) associated to a white node (taken cubic for simplicity), describing a superpotential term among D3-D3 fields $X_{ij}$ (labeled in black). The red arrows describe D7-branes connecting the relevant punctures, with D3-D7 sectors introducing flavors $q$, ${\tilde q}$ (labeled in black) and 73-33-37 couplings (\ref{supo0}) from the blue disks.\footnote{Moving the D7-brane 1-cycles pass the corresponding intersection between two zig-zag paths flips the orientation of blue disks. The presence of the 73-33-37 coupling is determined by the existence of the oriented disk.} In addition, there are D7-D7' fields $Y$, labeled in red. Their orientation, indicated by their subindices, has been determined by gauge invariance of the 37-77-73 couplings in \eref{W_73-33-37}. The oriented disks supporting the corresponding worldsheet instantons are shown in \fref{D7s_white_node_37-77-73}. Here we have used the discussion in \fref{D7s_white_node} for the chirality of 33 fields. 

\begin{figure}[h]
\begin{center}
\includegraphics[width=6cm]{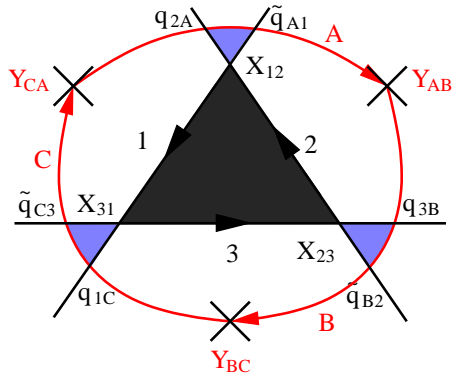}
\caption{D7-branes in the mirror for a set of punctures around a disk associated to a black node in the bipartite graph. Blue disks correspond to 73-33-37 couplings.}
\label{D7s_black_node}
\end{center}
\end{figure}

\fref{D7s_black_node} shows the analogous mirror configuration around a disk associated to a black node. While the orientation of the D7-branes is inverted, the chirality of D7-D7$'$ bifundamentals remains the same.

The previous discussion leads to an alternative prescription for determining the chirality of D7-D7$'$ fields in the simple case of D7-branes with a short embeddings sharing a puncture and a gauge group. Such a configuration is specified by three punctures. If the node enclosed by the triangle formed by the three punctures is white, the D7-D7$'$ field at the common puncture goes from the outgoing to the incoming D7-brane.  If it is black, the D7-D7$'$ field goes from the incoming to the outgoing D7-brane.

As already mentioned, the spectrum in the D7-D7$'$ sector is not protected against changes of the system which modify the behavior at infinity. A prominent example of such change corresponds to blowing up, i.e. resolution of the singularity. Although it would seem a local process, connecting different conical singularities by blowing up requires taking the blowup parameter to infinity (or equivalently, zooming infinitely into the residual singularity after partial blowup). Connecting singularities by partial blowup preserves the properties of fully localized fields, like those in D3-D3 or D3-D7 sectors, but it modifies the properties of the D7-D7' sector. Therefore we are led to the important conclusion that blowup cannot be exploited for relating the D7-D7$'$ spectra of different singularities.

\begin{figure}[h]
\begin{center}
\includegraphics[width=12.5cm]{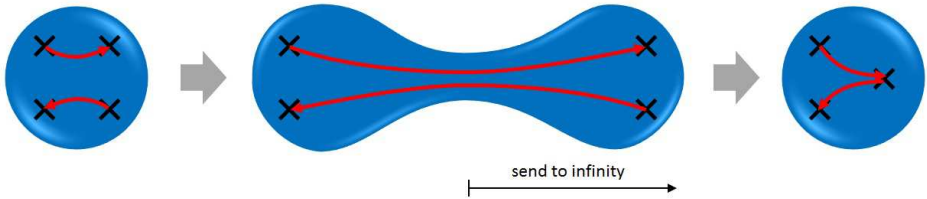}
\caption{\textit{Sudden change in the D7-D7$'$ spectrum under resolution.} Mirror picture of D7-branes in the conifold singularity corresponding to two paths without a common puncture. A finite blowup elongates the branes through a neck. In the limit of infinite blowup, the neck turns into a puncture, which is shared by the two D7-branes.}
\label{blowup-conifold}
\end{center}
\end{figure}

To illustrate this, consider the blowup of a conifold singularity to flat space $\mathbb{C}^3$. Recall that the conifold theory contains two gauge factors $U(N_1)\times U(N_2)$, and bifundamental matter multiplets $A_{12}$, $A'_{12}$, $B_{21}$, $B'_{21}$ \cite{Klebanov:1998hh}. We can introduce a D7 and a D7$'$-brane introducing flavors coupling to the D3-D3 bifundamentals $A_{12}$, $A'_{12}$, respectively. The D7 and D7'-branes do not share a puncture, so there are no relevant fields in the D7-D7' sector\footnote{This can also be seen in a T-dual picture \cite{Uranga:1998vf,Dasgupta:1998su,Park:1999ep}, with two (mutually rotated) NS5-branes on a circle, with D4-branes suspended between them; the D7 and D7'-branes turn into {\it half} D6-branes ending on the two different NS-branes \cite{Brodie:1997sz,Hanany:1997sa}, and hence with no massless open string stretched between them.}. The conifold singularity can be blown up to $\mathbb{C}^3$ by a vev for e.g. $B_{21}$; the fields $A_{12}$, $B'_{21
 }$ and $A'_{12}$ turn into the three adjoints $X$, $Y$, $Z$ of the ${\cal N}=4$ SYM theory of D3-branes in flat space. The D7 and D7'-branes now wrap the 4-cycles $X=0$ and $Z=0$ respectively, and there is a D7-D7' field $Y_{7\,7'}$ supported on the intersection 2-cycle, with a cubic 77$'$-7$'$3-37 coupling. The appearance of the new field $Y_{7\,7'}$ is an abrupt change during the infinite blowup connecting the systems. In the mirror picture, it is due to the appearance of a new puncture shared by the daughter D7-branes, see figure \ref{blowup-conifold}.

\bigskip

We conclude with another warning. In general, one can have D7-branes which share a puncture (at which they intersect with opposite orientation), but do not share a gauge group. Namely, the two D7-branes intersect the same zig-zag path in the dimer, but at distant locations not corresponding to consecutive edges. Such cases can be realized in orbifolds of $\mathbb{C}^3$, in which we can check there are no  D7-D7$'$ states. Hence we take the conservative view that such states are not present in general, i.e. our arguments do not suggest the existence of D7-D7$'$ states in such situation, and we do not invoke them in the construction of explicit examples in \sref{section_BFTs_from_branes}. Furthermore, as mentioned earlier, the D7-D7' spectrum in the examples in \sref{section_BFTs_from_branes} can also be determined by orbifold techniques in some parent theories. Note that, non-consecutive D7-branes sharing a puncture can turn into consecutive D7-branes sharing a puncture, by simply blowing up to remove all edges `separating' the D7-branes until they become consecutive. Our rule is that the consecutive case does produce D7-D7$'$ states and couplings, while the non-consecutive case does not; this is fully consistent, because of the above argument showing that blowing up can change the D7-D7$'$  spectrum.

\bigskip

\subsection{Long Embeddings}

\label{section_long_embeddings}

The understanding of D7-D7$'$ sectors allows to generalize the short embedding D7-branes in \sref{section_D7s_short_embeddings} to ones defined by long oriented paths in the dimer, to which from now on we refer to as {\it long embedding} D7-branes. Our discussion follows closely \cite{Forcella:2008au}, which contains the first systematic treatment of such embeddings (albeit for Euclidean 3-branes, although the analysis extends to flavor D7-branes with minimal modifications). We consider paths of the general form $\mathcal{O}_{i_0 i_n}=X_{i_0 i_1} X_{i_1 i_2} \ldots X_{i_{n-1}i_n}$, where consecutive fields not only share a common gauge group but also a puncture.\footnote{Open paths with consecutive fields not sharing a puncture (i.e. not in adjacent edges in the dimer) can be deformed to open paths satisfying this condition, by crossing over vertices in the dimer. Namely the corresponding operators differ by F-terms. In the discussion of dibaryons \cite{Forcella:2008au}, such operators are equivalent in the chiral ring, it is sufficient to keep one representative in each `homology' class. For the D7-brane case, different open paths should be regarded as different, since there may be obstructions to their equivalence (see discussion in \sref{sec:closed-loops}).}

\begin{figure}[h]
\begin{center}
\includegraphics[width=7cm]{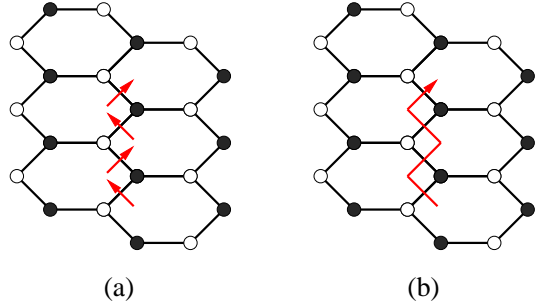}
\caption{\textit{Long embeddings from short ones.} a) The starting point is a set of D7-branes associated to sequence of consecutive short embeddings sharing faces of the dimer and punctures in $\Sigma$, equivalently zig-zag paths. b) By turning on non-zero vevs for D7-D7$'$ fields, the D7-branes combine into a single one with a long embedding.}
\label{embedding_long_path}
\end{center}
\end{figure}

In order to determine the gauge theory associated to such a long embedding, we first consider the theory of multiple `short embedding' D7-branes, D7$_{A_\mu}$ with $\mu=1,\ldots,n$, each one associated to a single chiral field $X_{i_{\mu-1} i_\mu}$ in the path. Each D7$_{A_\mu}$-brane gives rise to a pair of flavors $\tilde{q}_{A_\mu i_{\mu-1}}$ and $q_{i_\mu A_\mu}$, see Figure \sref{embedding_long_path}, which uses the arrow notation introduced in \sref{section_D7s_short_embeddings}. Furthermore, since consecutive D7s share a puncture, there are $Y_{A_\mu A_{\mu-1}}$ fields stretching between D7$_{A_\mu}$ and D7$_{A_{\mu-1}}$. They are invariant under the gauge symmetries but transform as bifundamentals of the corresponding global symmetry groups. 

Following \eref{supo0} and \eref{W_73-33-37}, the superpotential is given by
\begin{eqnarray}
W & = & \tilde{q}_{A_1i_0} X_{i_0 i_1} q_{i_1A_1}+\tilde{q}_{A_2i_1} X_{i_1 i_2} q_{i_2A_2}+\ldots+\tilde{q}_{A_{n}i_{n-1}} X_{i_{n-1} i_n} q_{i_nA_n} \nonumber \\
& - & q_{i_1A_1} Y_{A_1 A_2} \tilde{q}_{A_2i_1}-q_{i_2A_2} Y_{A_2 A_3} \tilde{q}_{A_3 i_2} - \ldots - q_{i_{n-1}A_{n-1}} Y_{A_{n-1}A_n} \tilde{q}_{A_ni_{n-1}}
\label{W_long_embedding_0}
\end{eqnarray}
For clarity, we have only presented the piece of the superpotential involving the flavor D7-branes, since completing the superpotential is straightforward in each given model. Throughout the rest of the paper, we will often take the same approach, hoping it does not lead to any confusion. 

Next, let us turn on non-zero expectation values for all the chiral fields $Y_{A_{\mu} A_{\mu+1}}$, recombining all D7$_{A_\mu}$-branes into a single D7-brane, labeled with $A$. The superpotential becomes
\begin{eqnarray}
W & = & \tilde{q}_{Ai_0} X_{i_0 i_1} q_{i_1A}+\tilde{q}_{Ai_1} X_{i_1 i_2} q_{i_2A}+\ldots+\tilde{q}_{Ai_{n-1}} X_{i_{n-1} i_n} q_{i_nA} \nonumber \\
& - & q_{i_1A}\tilde{q}_{Ai_1}-q_{i_2A} \tilde{q}_{Ai_2} - \ldots - q_{i_{n-1}A} \tilde{q}_{Ai_{n-1}} ,
\end{eqnarray}
where for simplicity we have set the vevs to 1. The quark-antiquark pairs for all intermediate faces in the dimer, $(q_{i_1A},\tilde{q}_{Ai_1}),\ldots, (q_{i_{n-1}A},\tilde{q}_{Ai_{n-1}})$, become massive. Integrating them out, we obtain the following superpotential
\beq
W = \tilde{q}_{Ai_0} \mathcal{O}_{i_0 i_n} q_{i_nA} .
\label{W_long_embedding_final}
\eeq
We thus obtain the straightforward generalization of \eref{supo0} to the case in which a single field is replaced by a path, where $\tilde{q}_{i_0}$ and $q_{i_n}$ are the surviving massless flavors at its endpoints. This reproduces the result advanced at the end of \sref{section_D7s_short_embeddings}.

Let us now consider two D7-branes given by paths $\mathcal{O}_{ij}$ and $\mathcal{O}'_{jk}$ of arbitrary length, sharing an intermediate face in the dimer $j$ and also a puncture. The previous analysis in terms of constituent embeddings makes it clear that in this case, as for short embeddings, there is a $Y_{\mathcal{O}\mathcal{O}'}$ field in the D7-D7$'$ sector. The full superpotential for this configuration takes the form

\beq
W= \tilde{q}_{i} \mathcal{O}_{i j} q_{j}+ \tilde{q}_{j} \mathcal{O}'_{j k} q_{k}+
\tilde{q}_j Y_{\mathcal{O}\mathcal{O}'} q_j .
\eeq

\bigskip
 
\subsection*{The Mirror}

\begin{figure}[h]
\begin{center}
\includegraphics[width=10.5 cm]{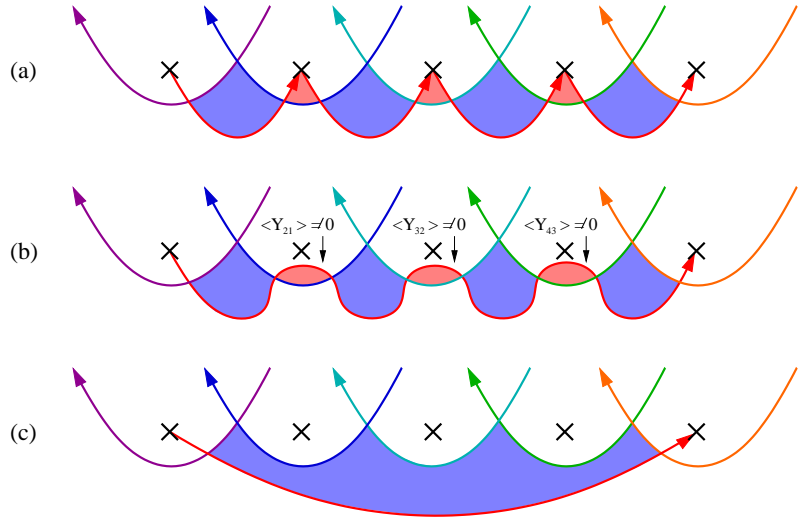}
\caption{\textit{Mirror perspective on long embeddings.} a) The starting configuration in terms of constituent short embedding D7-branes. b) Turning on vevs for D7-D7$'$ fields recombines the D7-branes into a single one and generates masses for the intermediate flavors. c) Final configuration after integrating out the massive fields. 
}
\label{long_embedding_mirror}
\end{center}
\end{figure}

\fref{long_embedding_mirror} shows the mirror counterpart of the previous discussion. The original configuration in terms of constituent branes is given in \fref{long_embedding_mirror}.a. All terms in the initial superpotential \eref{W_long_embedding_0} are clearly visible, and displayed as light blue and pink disks for 73-33-37 and 37-77-73 couplings, respectively.

Turning on vevs for the D7-D7$'$ fields $Y_{A_\mu A_{\mu+1}}$ fuses the D7-branes as depicted in \fref{long_embedding_mirror}.b. The pink disks contain two intersections on their boundary, corresponding to mass terms for the intermediate flavors. Integrating out these (now vector-like) fields corresponds to deforming the recombined D7-brane until the massive intersections disappear. The final result is given in \fref{long_embedding_mirror}.c, where the superpotential \eref{W_long_embedding_final} is manifest.

\bigskip
 
\subsection{General D7-Brane Configurations}

\label{section_general-branes}

In this section we investigate the gauge theory resulting from more complicated configurations of D7-branes.  Following our general approach, we do so by decomposing the desired configuration into simple constituents and studying how their recombination translates into field theoretic terms.
Rather than attempting an exhaustive classification, we present several classes illustrating the main physical phenomena. In particular, we focus on two interesting phenomena: First, in the presence of long embeddings, there can be D7-D7$'$ states even when the endpoints of the corresponding paths do not coincide. Secondly, these more general configurations can be used to produce superpotential couplings involving essentially arbitrary chains of D3-D3 and D7-D7$'$ fields joined by 37 fields, e.g. 37-77$'$-7$'$3-33$^n$, or 37-(77)$^m$-73-33.

\bigskip

\subsubsection{Osculating Long Embeddings and 37-77-73-33 Couplings}

\begin{figure}[h]
\begin{center}
\includegraphics[width=9cm]{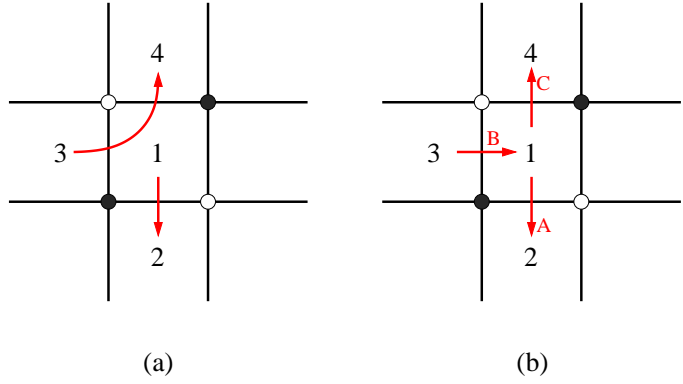}
\caption{a) The brane configuration we want to study. b) To do so, we first separate the length-2 embedding into constituent branes $B$ and $C$, which we later recombine by turning on a non-zero vev for $Y_{CB}$}
\label{dimer_new_coupling}
\end{center}
\end{figure}

Let us consider the explicit example shown in \fref{dimer_new_coupling}.a, involving two D7-branes with embeddings of length 1 and 2\footnote{Our discussion clearly applies to similar configurations, in which the valence of nodes or type of faces in the dimer are changed.}, described in the arrow notation introduced in \sref{section_D7s_short_embeddings}. Following our general strategy, we can start from \fref{dimer_new_coupling}.b and investigate what happens when the D7-branes B and C are recombined.

The spectrum and superpotential associated to \fref{dimer_new_coupling}.b can be determined using the rules for short embeddings and their interactions. 
The relevant part of the superpotential is given by 
\beqa
W=\tilde{q}_{A2} \, X_{21} \, q_{1A}+\tilde{q}_{B1} \, X_{13} \, q_{3B} + \tilde{q}_{C4} \, X_{41} \, q_{1C} - q_{1A} \, Y_{AB} \, \tilde{q}_{B1} -
q_{1C} \, Y_{CB} \, \tilde{q}_{B1} \, +\ldots\, , \nonumber \\
\label{W_quartic_0}
\eeqa
where $Y_{AB}$ and $Y_{CB}$ are D7-D7$'$ fields and  the notation is hopefully self-explanatory. The signs can be reabsorbed by field redefinitions, and are chosen for convenience. The dots indicate there are additional terms in the complete dimer model, but they are mere spectators in our argument, and are dropped herefrom.  In order to obtain the configuration in \fref{dimer_new_coupling}.a, we recombine branes $B$ and $C$ by turning on a non-zero vev for $Y_{CB}$. Without loss of generality, with set this vev to 1. The superpotential then becomes

\beq
W = \tilde{q}_{A2} \, X_{21} \, q_{1A}+\tilde{q}_{B1} \, X_{13} \, q_{3B} + \tilde{q}_{C4} \, X_{41} \, q_{1C} - q_{1A} \, Y_{AB} \, \tilde{q}_{B1} -
q_{1C} \,\tilde{q}_{B1} \, .
\label{W_length_2_1_mass_term}
\eeq
A mass term for the D3-D7 fields $q_{1C}$ and $\tilde{q}_{B1}$ is generated. They can be integrated out using their equations of motion
\begin{eqnarray}
q_{1C} & = & X_{13} \, q_{3B} - q_{1A} \, Y_{AB} \nonumber \\
\tilde{q}_{B1} & = & \tilde{q}_{C4} \, X_{41}  \, .
\end{eqnarray}
Plugging this back into \eref{W_length_2_1_mass_term}, we obtain
\beq
W=\tilde{q}_{A2} \, X_{21} \, q_{1A}+\tilde{q}_{C4} \, X_{41} \, X_{13} \, q_{3B} - \tilde{q}_{C4} \, X_{41} \, q_{1A} \, Y_{AB} \, .
\label{W_quartic_final}
\eeq
This exercise reveals two new features with respect to the simple examples considered in previous sections:

\begin{itemize}
\item The presence of a D7-D7$'$ field, $Y_{AB}$, which arises despite the endpoints of the embedding paths do not coincide. In the example at hand, $Y_{AB}$ is a field extending between the basic constituents, which survives the recombination of $B$ and $C$. Below we explain in detail how this is understood from the perspective of the mirror.
\item A new kind of superpotential term, $- \tilde{q}_{C4} \, X_{41} \, q_{1A} \, Y_{AB}$, whose structure is of the general form 37-D7-D7$'$-73-33: .
\end{itemize}

\bigskip

\subsection*{The Mirror}

The previous conclusions are beautifully captured by the geometry of the mirror as we explain below. The lessons learnt here extend to more involved setups.

\begin{figure}[h]
\begin{center}
\includegraphics[width=4.5cm]{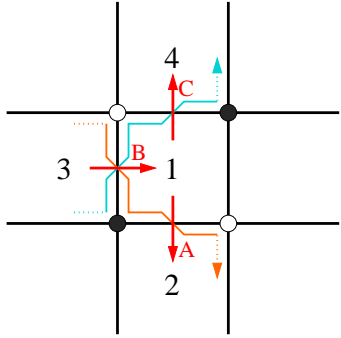}
\caption{The two zig-zag paths shared by the A-B and B-C pairs.}
\label{zig_zags_new_coupling}
\end{center}
\end{figure}

Before starting, let us make a simple observation about the system in \fref{dimer_new_coupling}.b. While the three constituent D7-branes share a same gauge group, given by face 1 of the dimer, they do not simultaneously coincide at a puncture in $\Sigma$. The puncture that is common to the A and B pair is different from the one that is shared by the B and C pair. This fact is clear from \fref{zig_zags_new_coupling} since, as explained in \sref{section_D3s_mirror}, punctures in $\Sigma$ correspond to zig-zag paths in the dimer model.

\begin{figure}[h]
\begin{center}
\includegraphics[width=8cm]{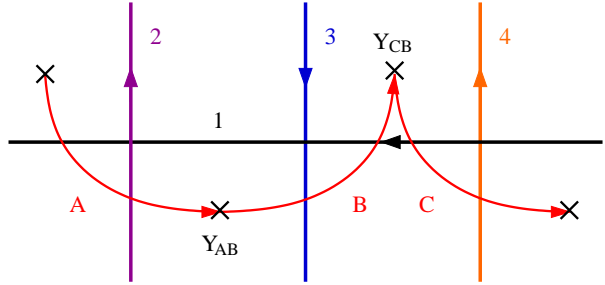}
\caption{Initial mirror configuration.}
\label{quartic_mirror_0}
\end{center}
\end{figure}

%
Following the discussion in \sref{section_D3s_mirror} and \sref{section_D7s_short_embeddings}, it is straightforward to determine that the relevant part of the mirror of \fref{dimer_new_coupling}.b is given by \fref{quartic_mirror_0}.\footnote{For simplicity, we omit the intersections between zig-zag paths associated to some of the edges in \fref{zig_zags_new_coupling}, which are not crucial for the discussion of the D7-branes.}

\begin{figure}[h]
\begin{center}
\includegraphics[width=15.5cm]{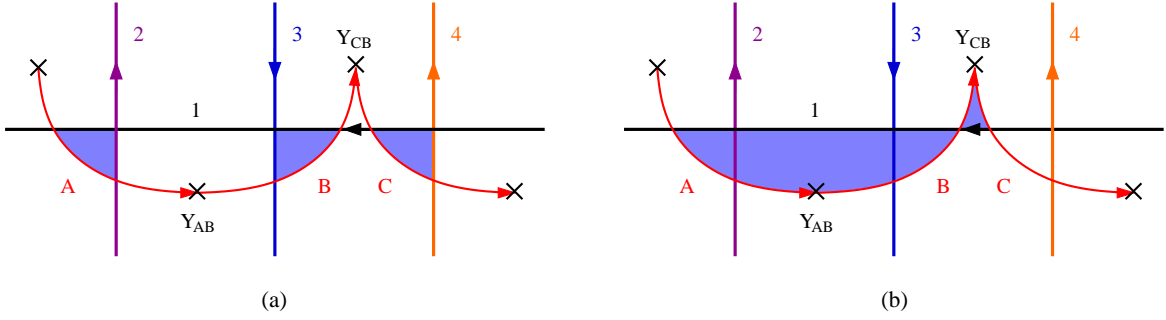}
\caption{\textit{Mirror description of the original superpotential.} All terms in \eref{W_quartic_0} are nicely captured by the mirror. a) $W=\tilde{q}_{A2} \, X_{21} \, q_{1A}+\tilde{q}_{B1} \, X_{13} \, q_{3B} + \tilde{q}_{C4} \, X_{41} \, q_{1C}$. b) $W=- q_{1A} \, Y_{AB} \, \tilde{q}_{B1} -
q_{1C} \, Y_{CB} \, \tilde{q}_{B1}$.}
\label{quartic_mirror_1}
\end{center}
\end{figure}

%
Indeed, as shown in \fref{quartic_mirror_1}, we can identify in the mirror all the terms in the superpotential \eref{W_quartic_0}.

\begin{figure}[h]
\begin{center}
\includegraphics[width=8cm]{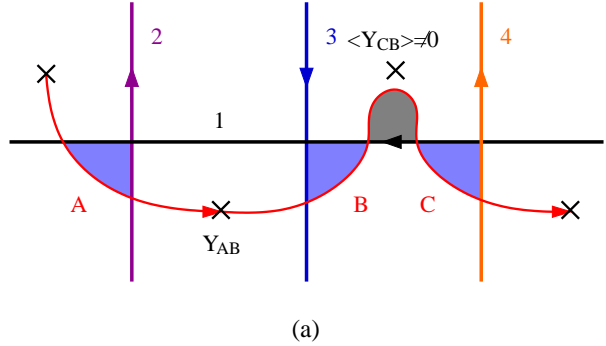}
\caption{\textit{Mirror configuration after B/C recombination.} We show several terms in the superpotential \eref{W_length_2_1_mass_term}. In blue, we indicate the terms $W=\tilde{q}_{A2} \, X_{21} \, q_{1A}+\tilde{q}_{B1} \, X_{13} \, q_{3B} + \tilde{q}_{C4} \, X_{41} \, q_{1C}$. The mass term $W=-
q_{1C} \, \tilde{q}_{B1}$ is shown in grey.}
\label{quartic_mirror_2}
\end{center}
\end{figure}

The D7-branes B and C are recombined by turning on a non-zero vev for $Y_{CB}$. The corresponding mirror configuration is given in \fref{quartic_mirror_2}, where we also indicate many of the superpotential terms in \eref{W_length_2_1_mass_term}.

\begin{figure}[h]
\begin{center}
\includegraphics[width=15.5cm]{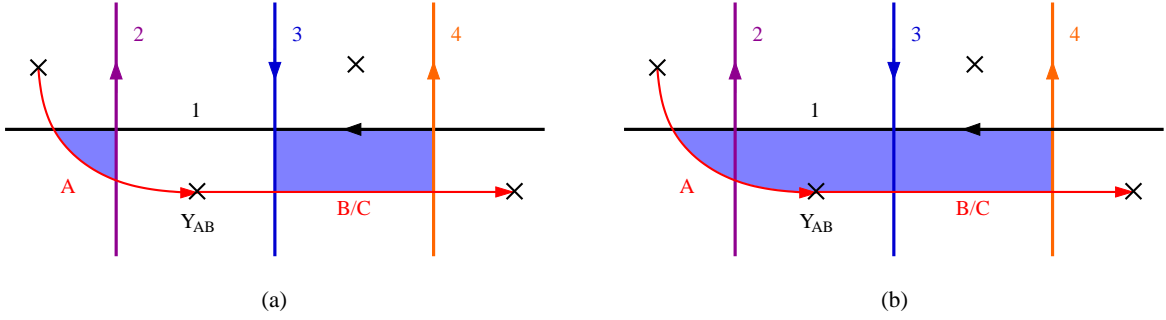}
\caption{\textit{Final mirror configuration.} The entire superpotential \eref{W_quartic_final} can be identified. a) $W=\tilde{q}_{A2} \, X_{21} \, q_{1A}+\tilde{q}_{C4} \, X_{41} \, X_{13} \, q_{3B}$. b) $W= - \tilde{q}_{C4} \, X_{41} \, q_{1A} \, Y_{AB}$.}
\label{quartic_mirror_3}
\end{center}
\end{figure}

As done in \sref{section_long_embeddings}, integrating out the massive flavors $q_1$ and $\tilde{q}_1$ corresponds to deforming the combined B/C brane such that its two intersections with the zig-zag path 1 that represent them disappears. \fref{quartic_mirror_3} gives the resulting configuration, where it is also shown how the entire superpotential \eref{W_quartic_final} arises.

Similar techniques can be used to study the general case of two D7's given by path of arbitrary length including consecutive edges of a face, dubbed {\em osculating} D7-branes. In the mirror,
the two branes intersect on a common puncture, supporting a D7-D7$'$ state. Its couplings can be determined by suitable decomposition into short constituent branes, which are subsequently recombined, as we continue showing in the following.

\bigskip

\subsubsection{General Couplings}

\begin{figure}[h]
\begin{center}
\includegraphics[width=2.5cm]{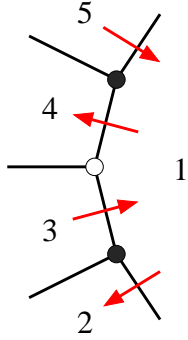}
\caption{\textit{A general configuration of flavor branes in the original dimer}. They all share a common dimer model face 1. In the mirror, each pair of consecutive D7s shares a puncture of $\Sigma$.}
\label{dimer_general_coupling}
\end{center}
\end{figure}

More general superpotential couplings can be similarly generated. Here we consider some further examples, which arise from the general configuration that is schematically shown in \fref{dimer_general_coupling} after turning on D7-D7$'$ vevs. This setup involves various D7-branes sharing a common dimer model face. In addition, each pair of consecutive D7s shares a zig-zag path, i.e. a puncture on $\Sigma$. We focus on the mirror description of this setup. Understanding the D7-brane recombination in terms of the bipartite graph or the field theory is also straightforward.

\bigskip

\subsection*{37-77-73-33$^2$ Couplings}

\begin{figure}[h]
\begin{center}
\includegraphics[width=11cm]{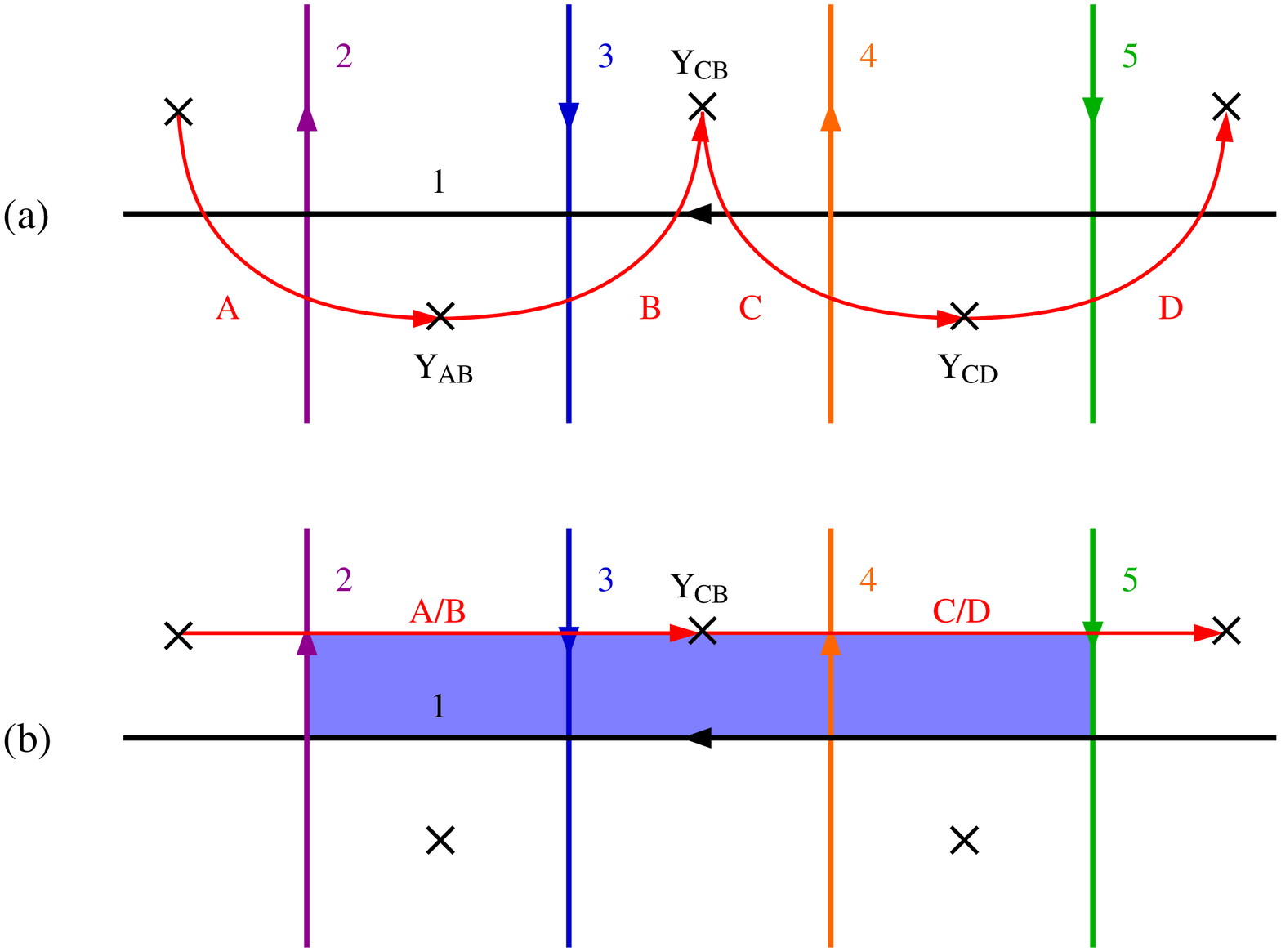}
\caption{a) The mirror of \fref{dimer_general_coupling}. b) Recombination after turning on vevs for $Y_{AB}$ and $Y_{CD}$.}
\label{mirror_37-77-73-33-33}
\end{center}
\end{figure}

\fref{mirror_37-77-73-33-33}.a shows the mirror of the general configuration in \fref{dimer_general_coupling}. Turning on vevs for $Y_{AB}$ and $Y_{CD}$, recombines the four D7-branes into two, as shown in \fref{mirror_37-77-73-33-33}.b. We observe the generation of a superpotential coupling of the form

\beq
W=q_{5D} \, Y_{CB} \, \tilde{q}_{B2} \, X_{21} \, X_{15}.
\eeq

\subsection*{37-77$^2$-73 Couplings}

\begin{figure}[h]
\begin{center}
\includegraphics[width=10cm]{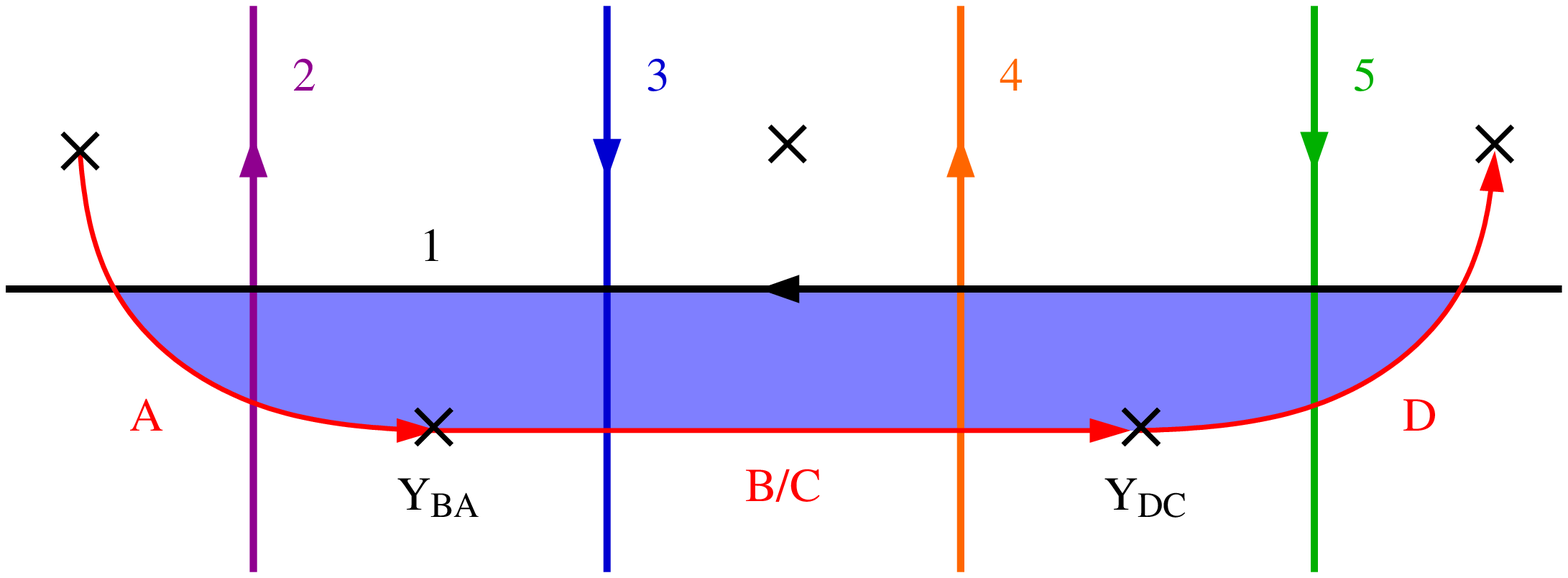}
\caption{The result of starting from \fref{mirror_37-77-73-33-33}.a and turning on a vev for $Y_{CB}$.}
\label{mirror_37-77-77-73}
\end{center}
\end{figure}

Starting again from the configuration shown in \fref{mirror_37-77-73-33-33}.a but turning a vev for $Y_{CB}$, one obtains \fref{mirror_37-77-77-73}. The following superpotential coupling is now generated

\beq
W=q_{1A} \, Y_{AB} \, Y_{CD} \, \tilde{q}_{D1} .
\eeq

\bigskip

The configurations we have just discussed are intended as simple illustrative examples. They can be easily generalized to give rise to more involved spectra and superpotential couplings. 

\bigskip

\subsection{Closed Loops and Obstructed Recombinations}
\label{sec:closed-loops}

We have seen that elementary D7-branes can be recombined, by giving vevs to the corresponding D7-D7$'$ fields, to produce couplings to open paths of bifundamentals. A natural question is whether it is possible to recombine D7-branes to form a closed loop in the dimer.\footnote{Here we mean turning on non-zero vevs for all D7-D7$'$ fields around the loop. When some D7-D7$'$ fields do not have vevs, the discussion is analogous to the case for open paths.} The answer reveals interesting physics, with two possibilities, depending on the homology of the loop on the dimer:

\medskip

\begin{itemize}

\item {\bf Homologically trivial loops:} there are couplings among the different D7-D7$'$ fields which prevent the simultaneous vevs for all D7-D7$'$ fields. 

\item {\bf Homologically non-trivial loops:} the simultaneous vevs are possible  and there is an induced coupling between a meson operator and the flavors. This is interpreted as the introduction of massive flavors, whose mass is controlled by the vev of the meson. Such couplings have appeared in \cite{Ganor:1996pe,Buican:2008qe}, in the context of D3-brane instantons.

\end{itemize}

\medskip

\begin{figure}[h]
\begin{center}
\includegraphics[width=11.5cm]{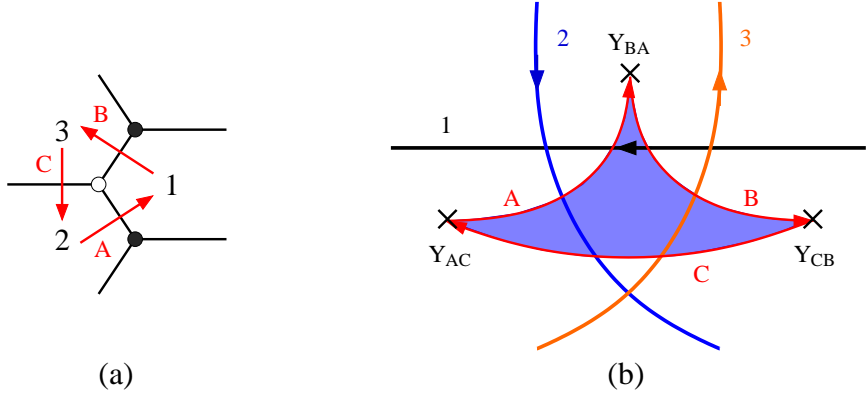}
\caption{\textit{General configuration leading to an obstruction for D7-D7$'$ vevs. } a) D7-branes forming a homologically non-trivial loop in the dimer. b) The mirror configuration, showing the disk responsible for the $Y_{BA}Y_{AC}Y_{CA}$ term in the superpotential, which prevents giving simultaneous vevs to all D7-D7$'$ fields.}
\label{dimer_mirror_obstruction_closed_loop}
\end{center}
\end{figure}

It is useful to discuss how these ideas work in an explicit example. Let us consider D3 and D7-branes on the $\mathbb{C}^3/\mathbb{Z}_3$ orbifold and 
introduce coordinates $x^i$ on $\mathbb{C}^3$. The D3-D3 bifundamentals are denoted $X^i_{a,a+1}$, and have superpotential
\beqa
W=\epsilon_{ijk} X^i_{a,a+1}X^j_{a+1,a+2}X^k_{a+2,a} .
\label{supo33-z3}
\eeqa
The mesonic moduli space is such that $x^i=X^i_{12}X^i_{23}X^i_{31}$ correspond to the coordinates of the D3-brane in the parent $\mathbb{C}^3$. We introduce D7$_k^a$-branes, defined by the 4-cycle $x^k=0$, with Chan-Paton phase determined by $a$. The corresponding flavors are denoted ${\tilde q}^k_{a,a+1}$, in the $(\fund^{7_k}_a,\antifund_{a+1})$ and $q^k_{a-1,a}$ in the $(\fund_{a-1},\antifund^{7_k}_a)$, and superpotential
\beqa
W_{3\,7}= {\tilde q}^k_{a,a+1} X^k_{a+1,a+2} q^k_{a+2,a} .
\label{supo37-z3}
\eeqa
In addition at the intersection between a D7$_i$ and D7$_j$ brane, there is\footnote{In toroidal orientifolds there are other fields, e.g. D7$_k$-D7$_k$ fields on the 4-cycle $z_k=0$. Following the general discussion, we rather focus on fields localized on curves at intersections of different 4-cycles, since their couplings to the 4d fields generalize to non-orbifold examples.} a $7_i7_j$ field $Y^k_{a,a+1}$, with $k\neq i\neq j\neq k$ and superpotential
\beqa
W= \epsilon_{ijk} Y^i_{a,a+1}Y^j_{a+1,a+2}Y^k_{a+2,a} + \epsilon_{ijk} q^i_{a,a+1} Y^j_{a+1,a+2} {\tilde q}^k_{a+2,a} .
\label{supo77-z3}
\eeqa

Consider the D7s associated to e.g. $X^1_{12}$, $X^2_{23}$ and $X^3_{31}$, which are all joined in a superpotential term \eref{supo33-z3}. This is the case where the D7s form a closed loop that is trivial in homology in the dimer. To form the D7 bound state, we need to introduce vevs for $Y^1_{12}$, $Y^2_{23}$ and $Y^3_{31}$. These vevs would, in turn, give masses to $q^2_{31}$, ${\tilde q}^3_{23}$, $q^3_{12}$, ${\tilde q}^1_{31}$, $q^1_{23}$ and ${\tilde q}^2_{12}$. But the first term in \eref{supo77-z3} prevents such simultaneous vevs. \fref{dimer_mirror_obstruction_closed_loop} shows an example of this situation for a general model, both from a dimer and mirror perspectives.

On the other hand, consider the D7s associated to e.g. $X^1_{12}$, $X^1_{23}$ and $X^1_{31}$, which form a closed loop that is not in the superpotential and is non-trivial in the dimer homology. It is easy to verify that in this case the superpotential \eref{supo77-z3} does not prevent making the corresponding flavors massive, thereby recombining the D7-brane. The interpretation is that different fractional D7s combine into a dynamical one which moves into the bulk. The mass of the flavors is controlled by the meson vev, which agrees with the fact that the flat direction is associated with the motion away from the singularity.

\bigskip

\section{BFTs from D-Branes}

\label{section_BFTs_from_branes}

In this section we illustrate how our ideas can be used for engineering BFTs. It is important to stress that the range of applicability of these tools is much wider and that they can be exploited to construct gauge theories which are not BFTs. 

We will focus on planar BFTs, i.e. theories determined by graphs on a disk, and non-planar BFTs associated to graphs with multiple boundaries on surfaces with zero curvature. In addition we will restrict to BFTs in which the symmetry groups associated to all faces, both internal and external, are abelian. Whether BFTs on Riemann surfaces with non-vanishing curvature can be constructed using D-branes is a very interesting question, but it is beyond the scope of this paper. With this goal in mind, the importance of having a general framework for understanding D7-branes is clear. In this setup they are responsible for the global symmetries, i.e. the external faces, of the BFTs. 

Part of the bipartite graph of the resulting BFT is inherited from the underlying dimer model. This graph is modified and extended in order to include the parts of the theory involving D7-branes.

We will focus on BFTs in which every external face is adjacent to two other ones. This type of faces correspond to D7-branes that share two punctures with other D7-branes.\footnote{Introducing D7-branes which share only one puncture with other ones, it is possible to construct theories closely related to the ones introduced in \cite{Xie:2012mr}. More concretely, such theories have the same quiver and superpotential of the ones in \cite{Xie:2012mr}, but the fields associated to external legs are 6d and hence not dynamical from a 4d perspective. These fields can be made dynamical by cutting-off the worldvolume of the D7-branes, which also results in the gauging of the symmetries associated to external faces. This behavior is analogous to the one exhibited by the theories constructed in \cite{Heckman:2012jh}.} Such D7-branes can be classified according to the occupancy of the two dimer faces they connect. They can be:

\medskip

\begin{itemize}
\item {\bf Occupied-Occupied (O-O):} they have two D3-D7 states and two D7-D7$'$ states. The corresponding external faces in the BFT are hence 4-sided. 
\item {\bf Occupied-Empty (O-E):} they have one D3-D7 state and two D7-D7$'$ states. The corresponding external faces in the BFT are 3-sided.
\end{itemize}

\medskip

\noindent Cancellation of twisted tadpoles sometimes requires the inclusion of additional D7-branes connecting two empty faces. We call them E-E branes. These branes play no role in the BFT but are necessary for consistency of the string theory construction.

Below we present several explicit examples. We start from models mainly involving the short embeddings of \sref{section_D7s_short_embeddings} and progressively move to more involved configurations. Our goal is by no means to provide a general classification of BFTs arising from D3-D7 configurations, but to illustrate the flexibility of our ideas and how they work in concrete configurations. 

Before concluding this section, we would like to mention that a string theory embedding for a class of quiver gauge theories associated to bipartite graphs on a disk, which are distinct but closely related to the BFTs discussed in this article, has been introduced in \cite{Heckman:2012jh}. The construction involves D5-branes and NS5-branes in Type IIB string theory. These branes share the four dimensions in which the gauge theory lives and have a non-trivial structure along two internal complex dimensions. D5-branes wrap an algebraic curve and NS5-branes wrap special Lagrangian submanifolds in the internal dimensions. This type of setup closely resembles brane tilings, which were introduced in \cite{Franco:2005rj}. One of the crucial differences is the fact that brane tilings are periodic along two directions.

\bigskip

\subsection{A Simple Example}

\begin{figure}[h]
 \centering
 \begin{tabular}[c]{ccc}
 \epsfig{file=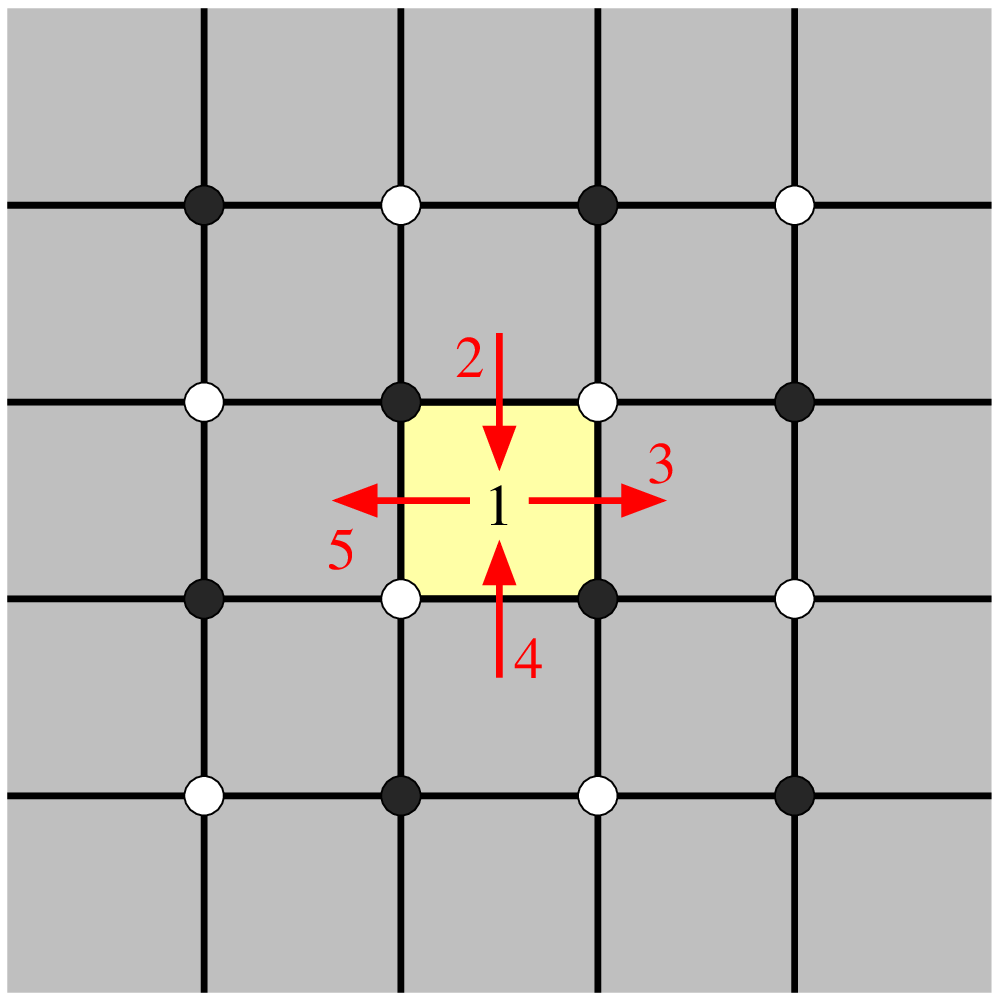,width=0.4\linewidth,clip=} & \ \ \ \ \ \ \ &
\epsfig{file=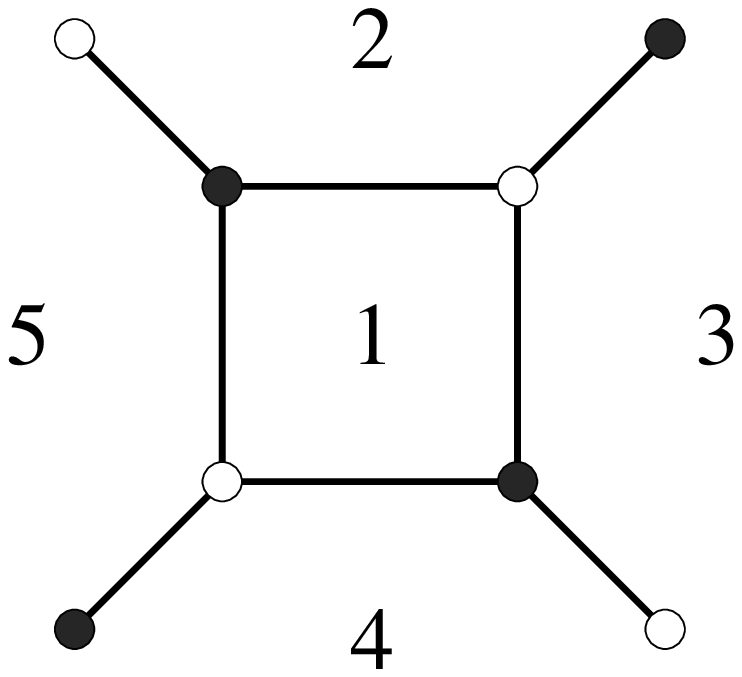,width=0.4\linewidth,clip=} \\ 
(a) & & (b)
 \end{tabular}
\caption{\textit{A simple example.} a) Brane configuration engineering the gauge theory under consideration. Occupied and empty faces in the underlying dimer model are shown in yellow and grey, respectively. Arrows indicate D7-branes. b) The resulting BFT.}
\label{BFT_G24}
\end{figure}

Let us begin our catalogue of explicit models by constructing a rather simple BFT. \fref{BFT_G24}.a summarizes the corresponding brane configuration. It contains a single fractional D3-brane, represented by the occupied yellow face in the figure. The underlying dimer model is a square lattice and can thus be embedded into a $\mathbb{Z}_N \times \mathbb{Z}_M$ orbifold of the conifold. The explicit values of $N$ and $M$ depend on the choice of unit cell, not shown in the figure, which is unimportant for our discussion. The configuration is completed with four D7-branes of O-E type, represented by red arrows.

As explained in \sref{section_anomalies_class_1}, tadpole cancellation corresponds to having an equal number of fundamental and antifundamental representations for all faces in the dimer, including those that are empty. It is straightforward to verify that \fref{BFT_G24}.a indeed satisfies this condition. For example, empty faces sharing an edge with face 1 have a flavor coming from this common edge, which is compensated by a flavor in the opposite direction from the corresponding D7-brane.

Let us now discuss the structure of the resulting BFT, whch is shown in \fref{BFT_G24}.b. The fractional D3-brane gives rise to an internal square face, since the D7-branes contribute two fundamentals and two antifundamentals to it. There is one external face for each of the D7-branes. Following \sref{section_77_sectors}, every consecutive pair of D7-branes supports a D7-D7$'$ field, i.e. an external leg in the BFT. In this example and the ones that follow, extending the constructions to non-abelian BFTs simply amounts to including multiple D-branes in each stack.

\bigskip

\subsection{Class 1}

\begin{figure}[h]
 \centering
 \begin{tabular}[c]{ccc}
 \epsfig{file=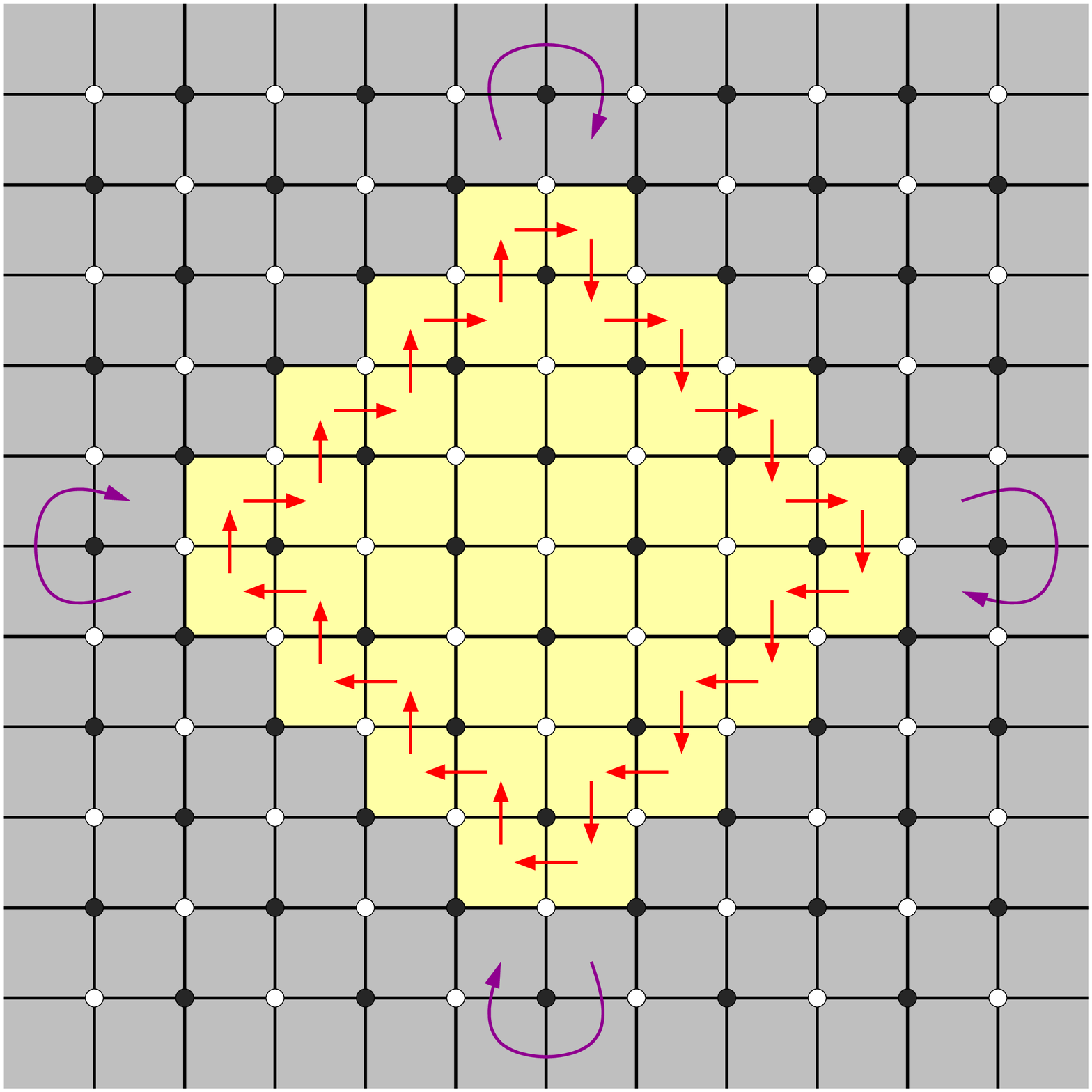,width=0.4\linewidth,clip=} & \ \ \ \ &
\epsfig{file=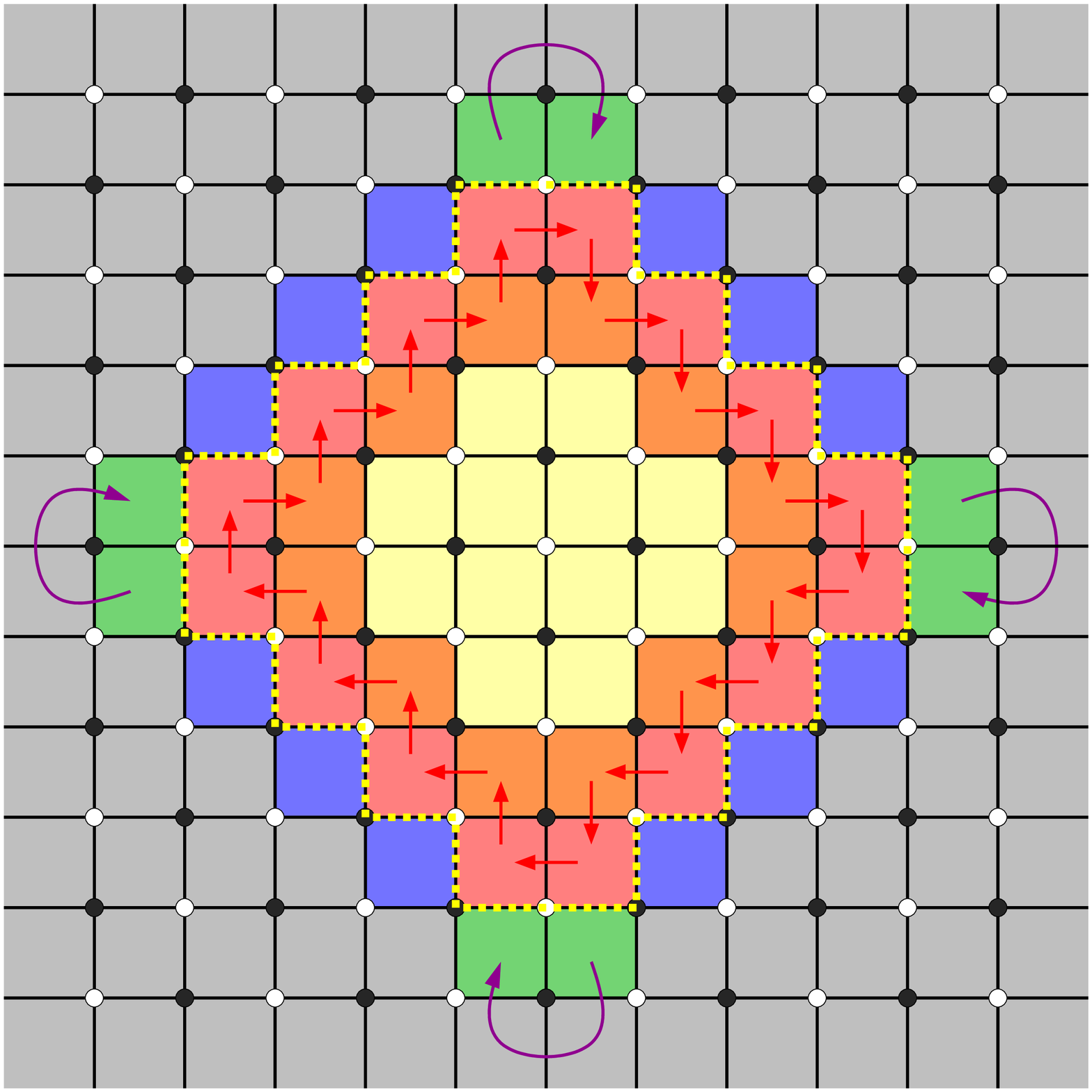,width=0.4\linewidth,clip=} \\ 
(a) & & (b)
 \end{tabular}
\caption{\textit{Brane configuration for Class 1 models.} a) Occupied and empty faces in the underlying dimer model are shown in yellow and grey, respectively. Arrows indicate D7-branes. b) A refined color coding for the faces in the dimer. The boundary between occupied and empty faces is indicated with a dashed yellow line.}
\label{dimer_classA} 
\end{figure}

We will now construct an infinite family of models, which corresponds to the D-brane configuration shown in \fref{dimer_classA}.a. The figure shows one representative in this class of models. Occupied faces define a rectangular area which can have arbitrary lengths in its two directions. Once again, these theories are based on square dimer models and can be embedded in $\mathbb{Z}_N \times \mathbb{Z}_M$ orbifolds of the conifold. In the general case, one simply takes $N$ and $M$ sufficiently large to accommodate the desired theory.  

All external faces of the BFT arise from D7-branes of O-O type. Each pair of consecutive O-O branes, which give rise to the boundary of the BFT, shares a puncture on $\Sigma$. This is because they sit on adjacent edges of faces of the dimer and hence intersect a common zig-zag path. As explained later, the additional E-E D7-branes represented by purple arrows are necessary for twisted tadpole cancellation.

\bigskip

\subsubsection*{Tadpole Cancellation}

\label{section_anomalies_class_1}

For discussing tadpole cancellation and the resulting BFT, which is presented in the following section, it is convenient to refine the color coding of faces in the dimer as shown in \fref{dimer_classA}.b. In these models there are three types of occupied faces (yellow, orange and pink) and three types of empty faces (grey, blue and green).

As explained in \sref{sec:anomaly}, tadpole cancellation is equivalent to anomaly cancellation for all faces in the dimer, including empty ones. We now go over each type of occupied and empty faces and discuss how anomalies are cancelled. For simplicity we consider the abelian case, in which all occupied faces have rank 1; the discussion for more general (but equal) ranks is similar, by simply increasing the number of D7-branes accordingly.

Let us first consider occupied faces, for which we have:

\begin{itemize}
\item {\bf Yellow:} they sit at the `bulk' of the graph. They do not have any D3-D7 flavor. Cancellation of anomalies proceeds as in the original dimer.
\item {\bf Orange:} the D3-D3 sector is given by the four edges of the original dimer, which correspond to an equal number of fundamental and antifundamental fields. In addition, there is one fundamental and one antifundamental in the D3-D7 sector.
\item {\bf Pink:} there is one fundamental-antifundamental pair coming from the D3-D3 sector and another one from the D3-D7 sector.
\end{itemize}
In all cases, the numbers of fundamental and antifundamental fields are equal for all internal faces.

Let us now consider empty faces. We discuss them in terms of the chiral fields that would exist if these faces were occupied. For the three different types of them, we have:

\begin{itemize}
\item {\bf Grey:} there are no fields charged under them.
\item {\bf Blue:} they have two D3-D3 states with opposite orientations.
\item {\bf Green:} they have a single edge in common with occupied faces, giving a single fundamental or antifundamental D3-D3 field, which would make the faces anomalous in the absence of other contributions. These anomalies can be cancelled by introducing the four D7-branes represented by purple arrows. Their only purpose is tadpole cancellation and they are decoupled from the BFT.

\end{itemize} 

\bigskip

\subsection*{The BFTs}

The matter content of this class of theories has been discussed in detail in the previous section. Their superpotential follows from the rules in \sref{section_D7s_short_embeddings}. The BFTs thus have the following structure.

\bigskip
\medskip

\paragraph{Internal Faces:} internal faces in the BFT, i.e. gauge symmetries, arise from occupied faces in the underlying dimer model. Yellow faces remain square. Orange faces have four D3-D3 edges plus two additional ones coming from D3-D7 states, becoming hexagons. Finally, pink faces have two D3-D3 edges plus two D3-D7 ones, and are hence squares.

\bigskip

\paragraph{External Faces:} all external faces in this class of models have the same structure. They have four edges, two from D7-D7$'$ and two from D3-D7 fields. Since they have an even number of edges, all global symmetries are anomaly free. The edges around any external face participate in three superpotential nodes, one of type \eref{W_73-33-37} and two of type \eref{supo0}.

\bigskip
\medskip

\begin{figure}[h]
\begin{center}
\includegraphics[width=8cm]{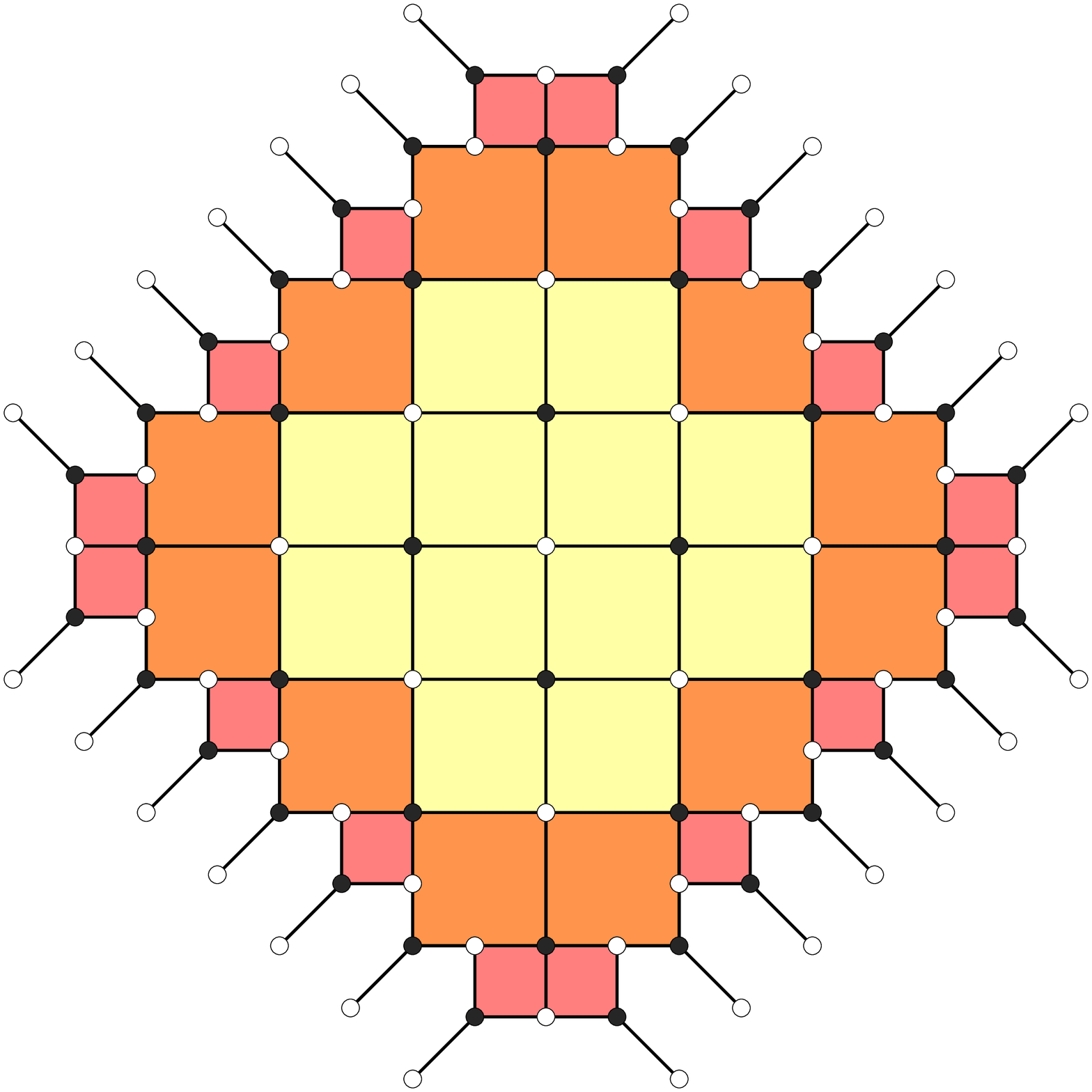}
\caption{\textit{Class 1 BFTs.} The colors in \fref{dimer_classA} are used to indicate the origin of internal faces.}
\label{BFT_classA}
\end{center}
\end{figure}

\noindent The bipartite graph defining the resulting BFT is shown in \fref{BFT_classA}. 
All external nodes have the same color. Interestingly, modulo the fact that the chiral fields associated to external legs are non-dynamical from a 4d viewpoint, these theories are indeed in the special sub-class considered in \cite{Xie:2012mr}. In order to obtain more general field theories, with both white and black external nodes, it is necessary to include O-E D7-branes. We do so in the examples that follow.

\bigskip

\bigskip

\subsection{Class 2}

\begin{figure}[h]
 \centering
 \begin{tabular}[c]{ccc}
 \epsfig{file=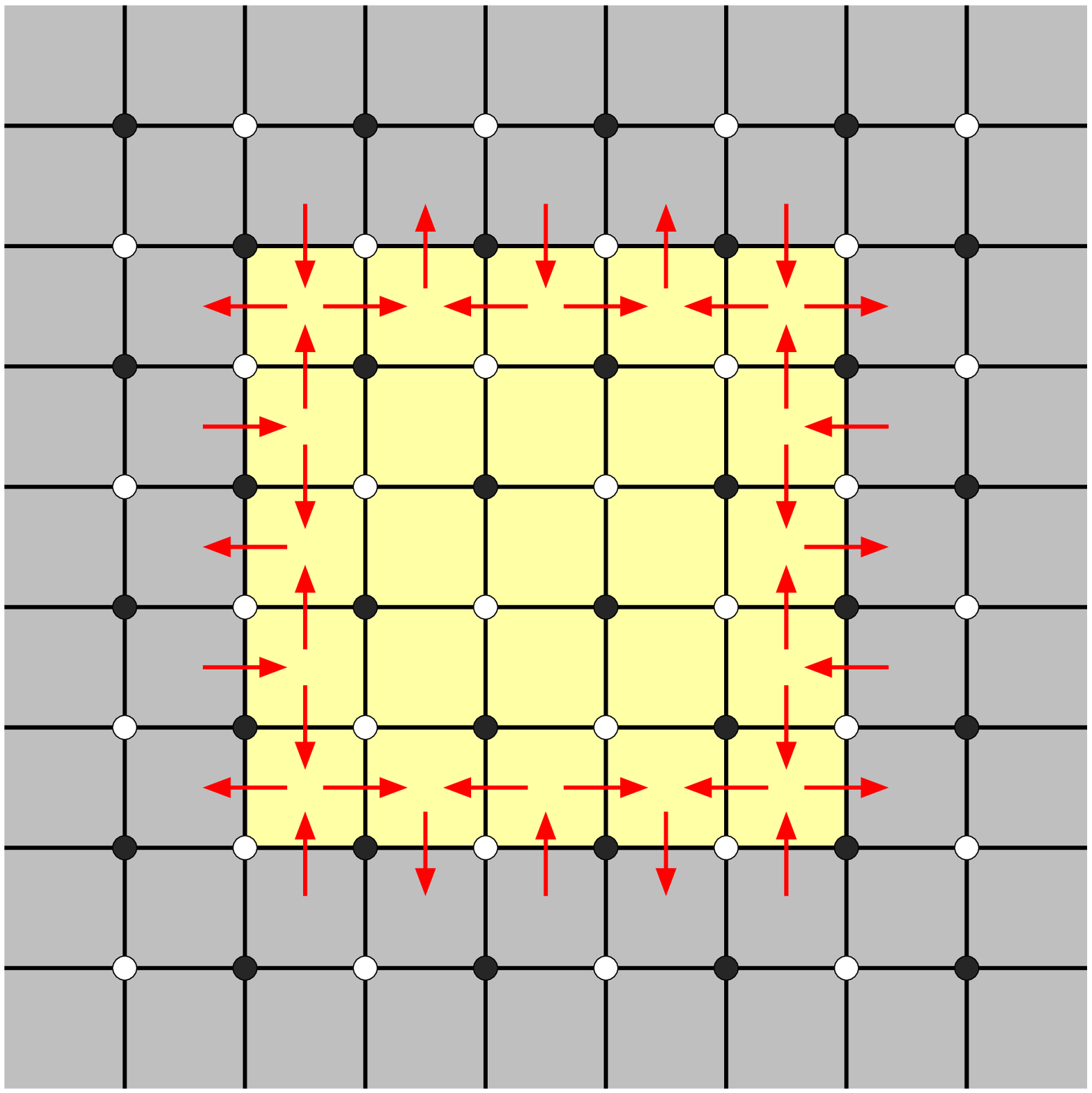,width=0.4\linewidth,clip=} & \ \ \ \ &
\epsfig{file=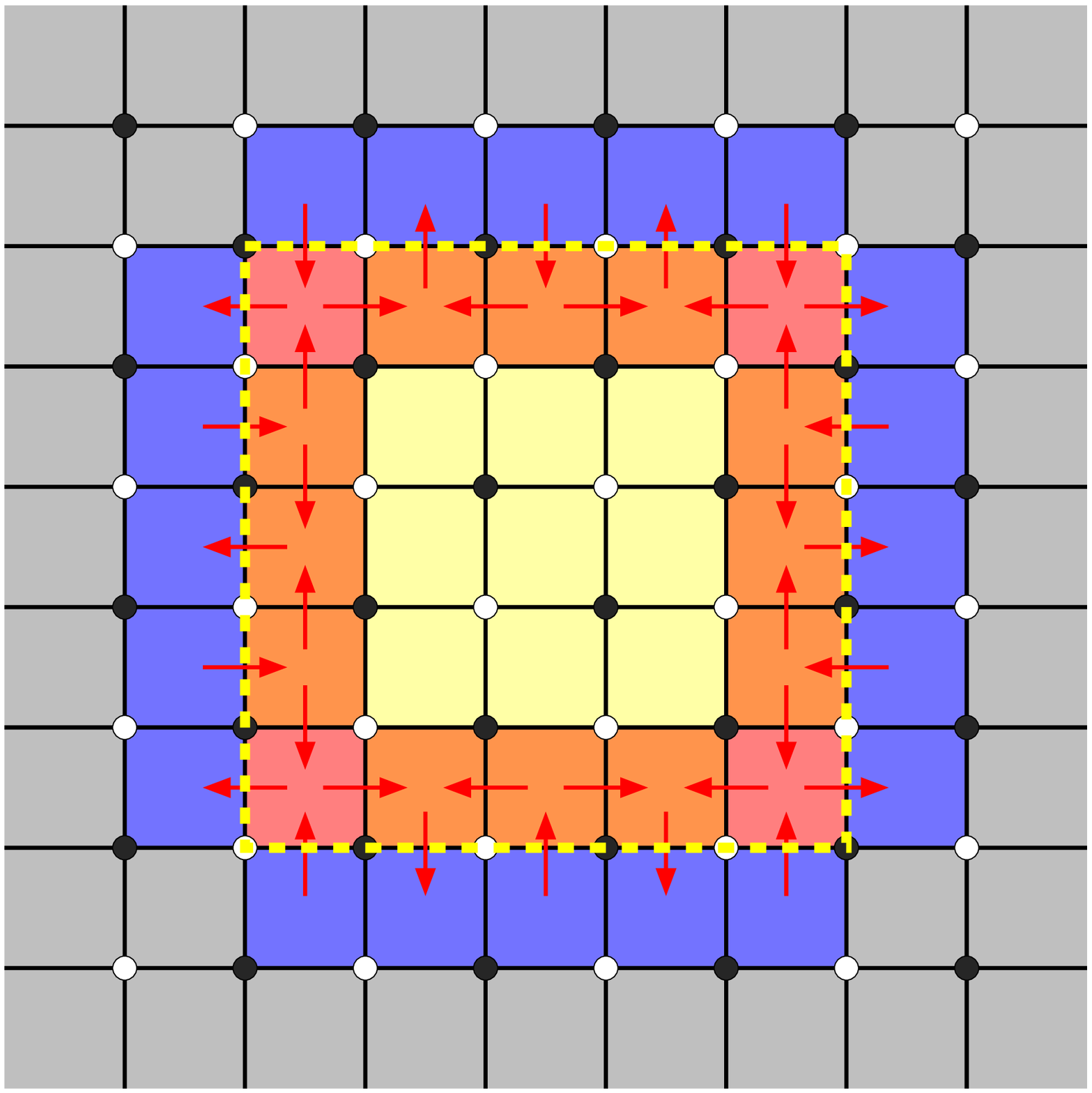,width=0.4\linewidth,clip=} \\ 
(a) & & (b)
 \end{tabular}
\caption{\textit{Brane configuration for Class 2 models.} a) Occupied and empty faces in the underlying dimer model are shown in yellow and grey, respectively. Arrows indicate D7-branes. b) A refined color coding for the faces in the dimer. The boundary between occupied and empty faces is indicated with a dashed yellow line.}
\label{dimer_classB} 
\end{figure}

We now introduce a second class of theories. Since analysis parallels the one for Class 1, we can be more schematic. These theories are also based on $\mathbb{Z}_N \times \mathbb{Z}_M$ orbifolds of the conifold and the basic configuration of D-branes is shown in \fref{dimer_classB}.a. The dimension of the rectangular occupied area in the dimer is arbitrary. In this case, both O-O and O-E D7-branes are included. Unlike Class 1, E-E D7-branes are not necessary for tadpole cancellation.

\subsection*{Tadpole Cancellation}

\label{section_anomalies_B}

As before, it is useful to refine the color coding of dimer faces, as shown in \fref{dimer_classB}.2. For the occupied faces, we have:

\begin{itemize}
\item {\bf Yellow:} they sit at the bulk of the graph. They do not have any D3-D7 flavor. Cancellation of anomalies proceeds as in the original dimer.
\item {\bf Orange:} they contain only three D3-D3 edges of the original dimer, since the remaining one is in contact with an empty face. In addition, there are three D3-D7 states with appropriate orientations to cancel the anomaly.\footnote{For practical purposes, it is useful to notice that whenever there is a D7-brane whose embedding corresponds to a single edge in the dimer between two occupied faces (i.e. a sub-class of the O-O branes), then the corresponding D3-D3 and D3-D7 pair has a zero net contribution to the anomaly.}
\item {\bf Pink:} these faces have two D3-D3 and four D3-D7 states.
\end{itemize}
The numbers of fundamental and antifundamental fields are equal for all internal faces. There are only two different types of empty faces:

\begin{itemize}
\item {\bf Grey:} there are no fields charged under them.
\item {\bf Blue:} they have one D3-D3 state and one D3-D7 state with opposite contribution to the anomaly.
\end{itemize} 

\bigskip

\subsection*{The BFTs}

The internal and external faces in the bipartite graph of the BFT are described below.

\bigskip

\paragraph{Internal Faces:} internal faces in the BFT correspond to occupied faces in the underlying dimer model. Yellow faces remain square. Orange faces have three D3-D3 edges plus three additional ones coming from D3-D7 states. They become hexagons in which three edges are in contact with external faces. Pink faces have two D3-D3 edges plus four D3-D7 ones, becoming hexagons.

\bigskip

\paragraph{External Faces:} this class of theories has external faces with three and four edges, corresponding to O-E and O-O D7-branes, respectively.

\bigskip
\medskip

\begin{figure}[h]
\begin{center}
\includegraphics[width=7cm]{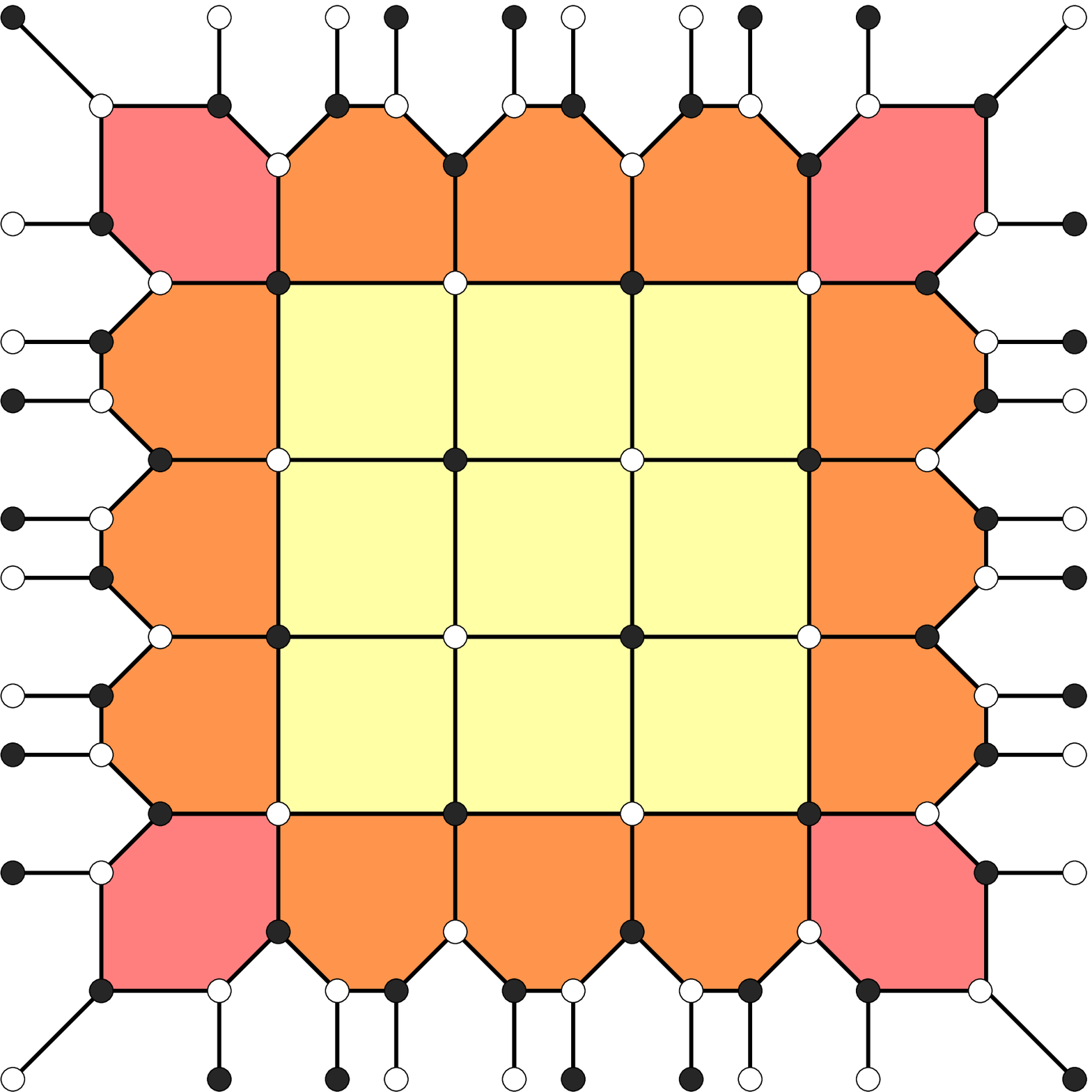}
\caption{\textit{Class 2 BFTs.} The colors in \fref{dimer_classB} are used to indicate the origin of internal faces.}
\label{BFT_classB}
\end{center}
\end{figure}

The resulting BFT is shown in \fref{BFT_classB}. The inclusion of O-E D7-branes results in external nodes of two colors. 

\bigskip

\subsection{Models Involving Long Embeddings}

\begin{figure}[h]
 \centering
 \begin{tabular}[c]{ccc}
 \epsfig{file=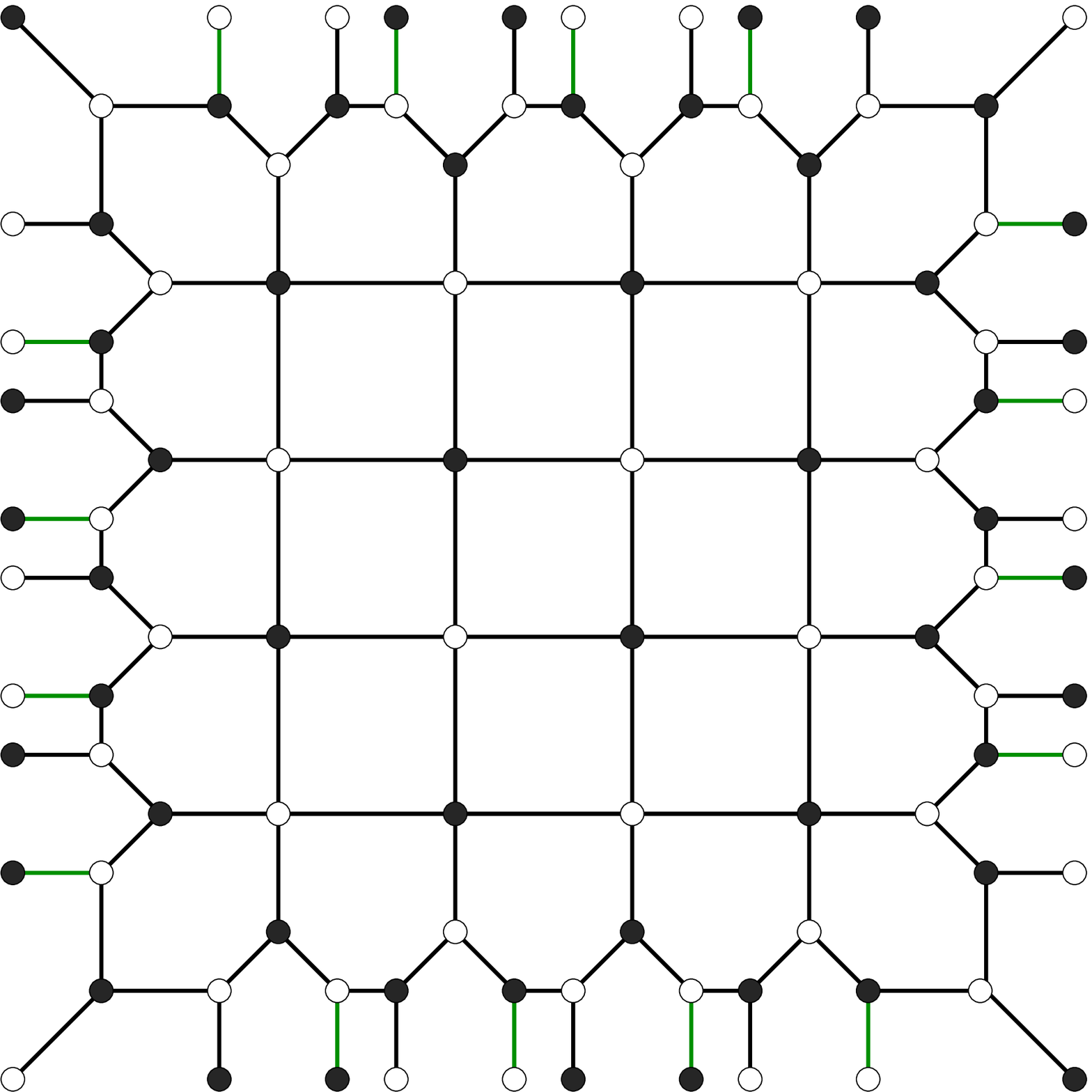,width=0.4\linewidth,clip=} & \ \ \ \ \ &
\epsfig{file=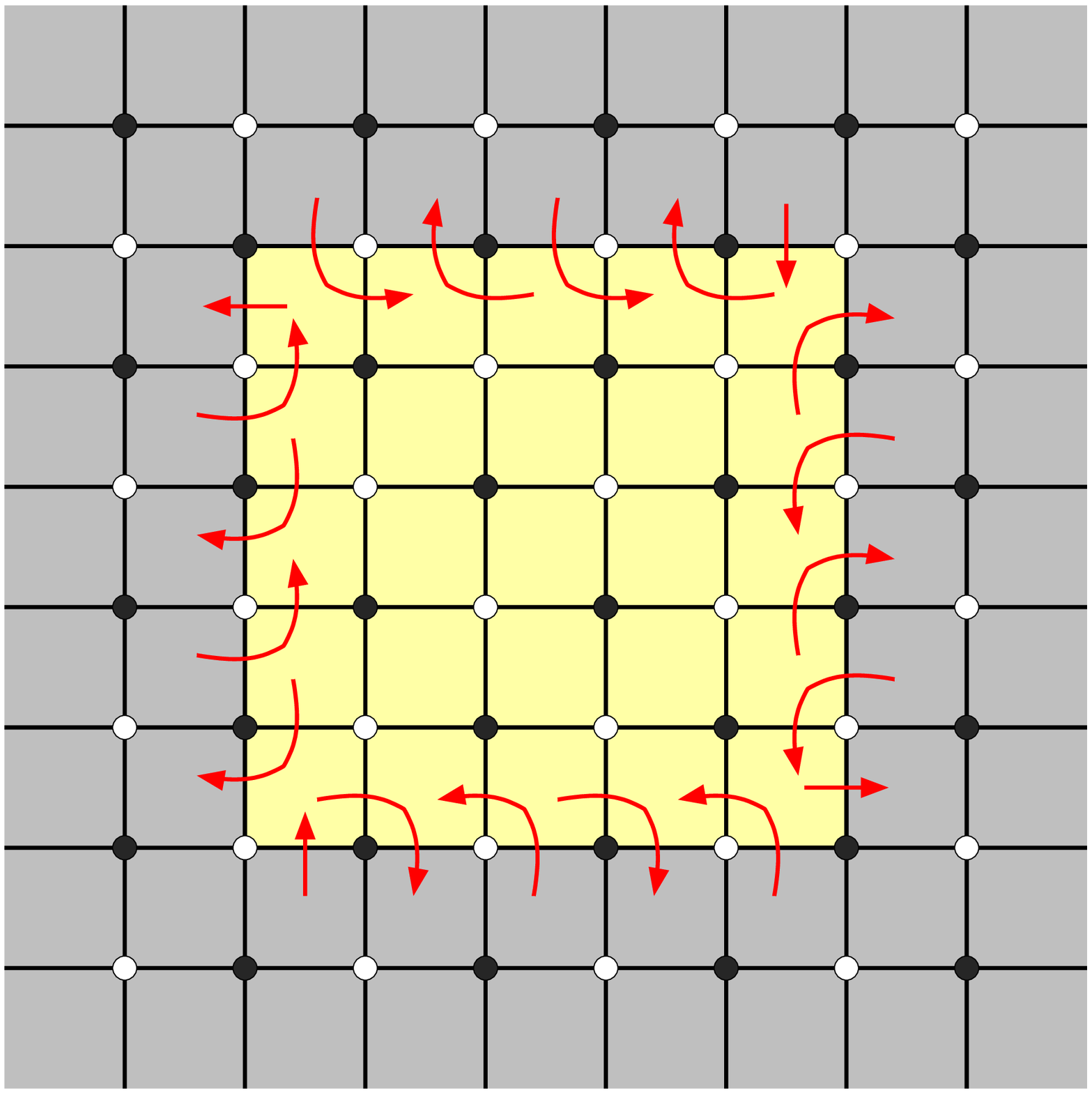,width=0.4\linewidth,clip=} \\ 
(a) & & (b)
 \end{tabular}
\caption{\textit{Class 3 BFTs.} a) A general representative of the Class 2 of BFTs. We will turn on vevs for the green edges. b) The brane configuration obtained after recombination.}
\label{BFT_classB_higgsed} \label{dimer_classB_higgsed} 
\end{figure}

We now present BFTs engineered using D7-branes with long embeddings, which we will call Class 3. They can be constructed by starting from BFTs based on short D7-branes embeddings and combining them by turning on D7-D7$'$ vevs. Let us start from Class 2 models and turn on vevs for the fields associated to the green edges in \fref{BFT_classB_higgsed}.a. The D-brane configuration is obtained from \fref{dimer_classB} by recombination and is given in \fref{dimer_classB_higgsed}.b.

\fref{BFT_classB_higgsed} shows the resulting BFT, which is obtained after integrating out massive fields. This result is in agreement with direct application of the rules in 
\sref{section_D7s_short_embeddings} and \sref{sec:d7s-full} to \fref{dimer_classB_higgsed}.
%
\begin{figure}[h]
\begin{center}
\includegraphics[width=6cm]{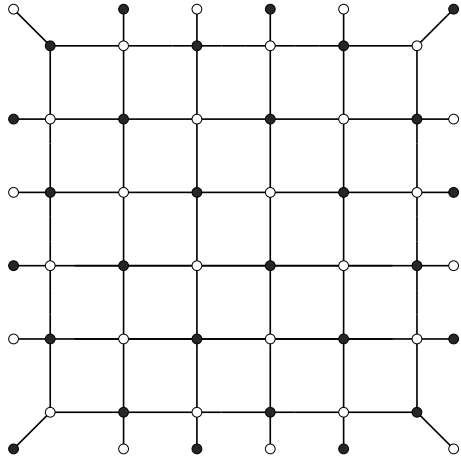}
\caption{A BFT in Class 3.}
\label{BFT_classB_higgsed}
\end{center}
\end{figure}

\bigskip

\subsection{Beyond the Disk}

\label{section_non_planar_BFTs}

\begin{figure}[h]
 \centering
 \begin{tabular}[c]{ccc}
 \epsfig{file=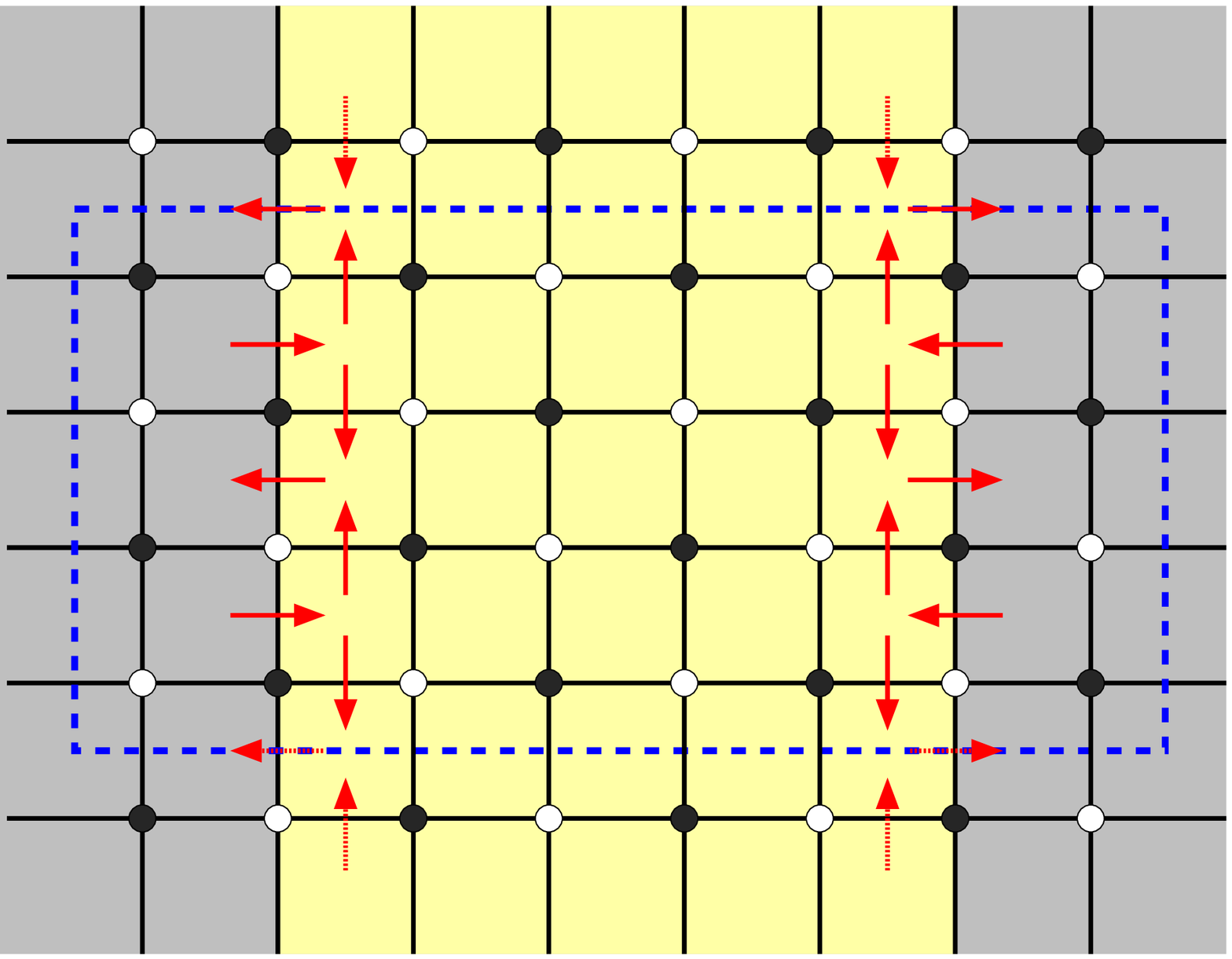,width=0.4\linewidth,clip=} & \ \ \ \ \ &
\epsfig{file=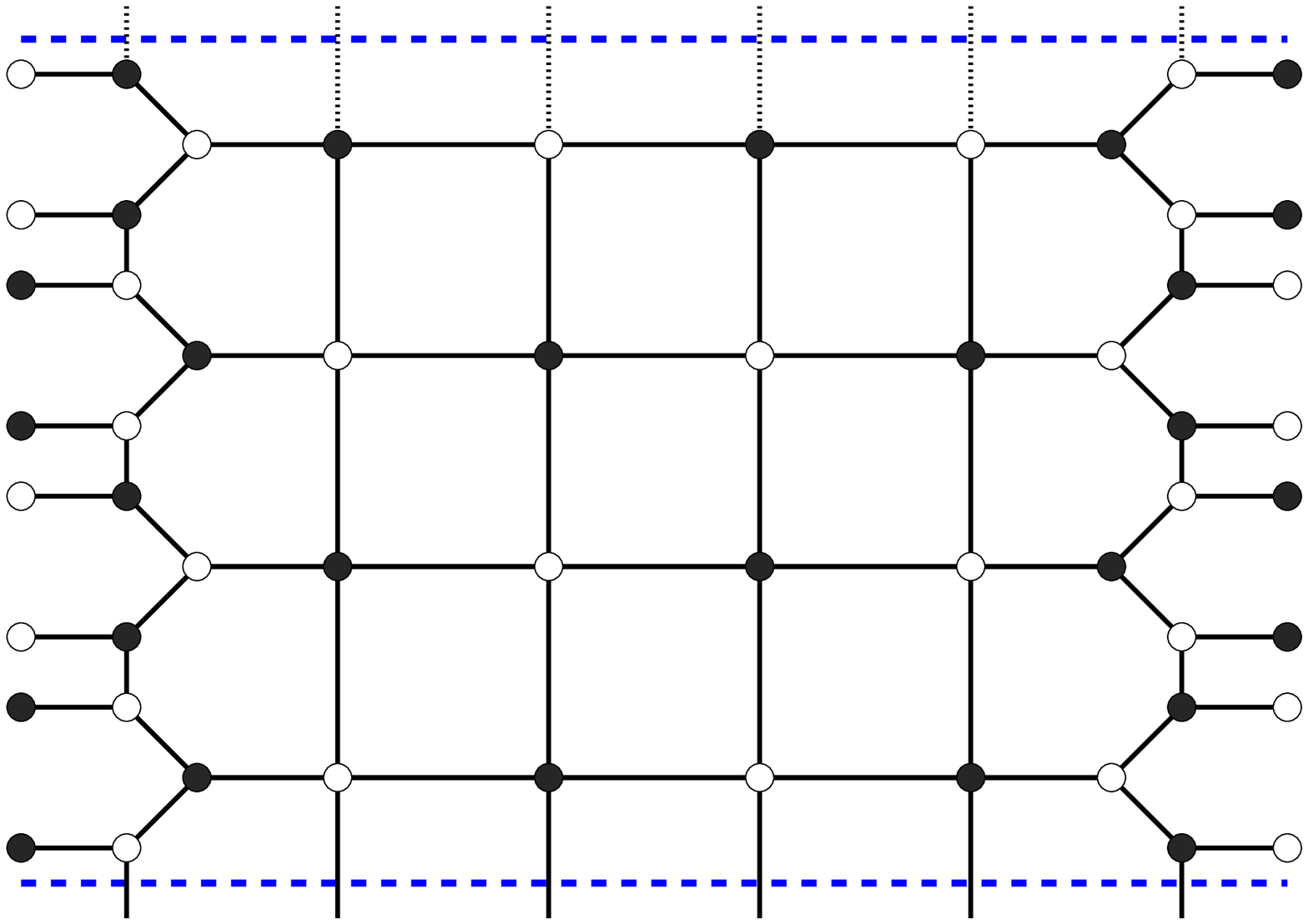,width=0.4\linewidth,clip=} \\ 
(a) & & (b)
 \end{tabular}
\caption{\textit{A non-planar BFT.} a) A configuration of D3 and D7-branes. The unit cell of the dimer model is shown in blue. b) The resulting BFT is non-planar and lives on a cylinder. The bipartite graph should be identified along the dashed blue lines.}
\label{BFT_cylinder} 
\end{figure}

The tools we have introduced allow the construction of non-planar BFTs. For example, let us consider the configuration shown in \fref{BFT_cylinder}.a. The unit cell is explicitly indicated by blue dashed lines and corresponds to a $\mathbb{Z}_2\times \mathbb{Z}_8$ orbifold of the conifold. This arrangement is closely related to the one for Class 2 theories. The main difference is that occupied faces wrap entirely one of the compact directions of the 2-torus. \fref{BFT_cylinder}.b shows the resulting BFT, which lives on a cylinder.

It is straightforward to exploit our ideas for constructing non-planar BFTs with multiple boundaries.

\bigskip

\section{Conclusions}

\label{section_conclusions}

We developed a comprehensive framework for determining the gauge theories arising on general configurations of D-branes over toric CY 3-folds in Type IIB string theory. The main contribution of our work is a significant extension of the understanding of flavor D7-branes, i.e. D7-branes wrapping non-compact 4-cycles, which was previously primarily restricted to simple embeddings and/or orbifold geometries. Our approach combines dimer models, mirror symmetry and the construction of general embeddings by recombination of elementary D7-branes.

In order to illustrate our ideas, we discussed in detail how to engineer a large sub-class of the BFTs introduced in \cite{Franco:2012mm}, corresponding to graphs with vanishing curvature. Several examples were presented, including infinite families of models and non-planar theories. We expect the D-brane realization will shed further light on the physics of BFTs. 

The range of applicability of the methods introduced in this article is far more general. In fact, the introduction of D7-branes in D3-brane systems has proven useful in a variety of contexts. For instance, the construction of local embeddings of phenomenological Particle Physics models or, replacing D7-branes by Euclidean D3-branes, the study of stringy non-perturbative contributions to quantum field theories. The latter, either with Euclidean D3-branes, or via non-perturbative effects in the corresponding D7-branes, has found applications to the questions of brane inflation \cite{Baumann:2006th,Baumann:2007np}, and of the generation of Yukawa couplings \cite{Marchesano:2009rz,Font:2012wq}. Finally, our construction of more general D7-branes may have application in the context of introducing flavours in gauge/gravity dual pairs, generalizing some of the existing constructions, e.g. \cite{Ouyang:2003df,Kuperstein:2004hy} for the conifold.

In \cite{ArkaniHamed:2001ie}, the 6d (2,0) and little string theories were deconstructed in terms of 4d quivers. These quivers correspond to $\mathcal{N}=2$ and $\mathcal{N}=1$ orbifolds of $\mathbb{C}^3$ which, interestingly, are BFTs. It is natural to conjecture that BFTs also deconstruct 6d theories on more general Riemann surfaces, which might include the $\mathcal{N}=1$ theories considered in \cite{Bah:2011vv,Bah:2012dg}. If this is the case, the D-brane realization of some of the models introduced in this paper may provide a useful tool for establishing the correspondence, as it did in \cite{ArkaniHamed:2001ie}.

\bigskip

\section*{Acknowledgments}

S. F. would like to thank D. Galloni, A. Mariotti and R.-K. Seong for useful discussions and collaboration on related problems. The work of S. F. is supported by the U.K. Science and Technology Facilities Council (STFC). The work of A. U. is supported by the Spanish Ministry of Economy and Competitiveness under grants FPA2010-20807, FPA2012-32828, Consolider-CPAN (CSD2007-00042),  the grants  SEV-2012-0249 of the Centro de Excelencia Severo Ochoa Programme,  HEPHACOS-S2009/ESP1473 from the C.A. de Madrid, SPLE Advanced Grant under contract ERC-2012-ADG$\_$20120216-320421 and the contract ``UNILHC" PITN-GA-2009-237920 of the European Commission. 

\bigskip
\bigskip



\begin{thebibliography}{10}

\bibitem{Maldacena:1997re} 
  J.~M.~Maldacena,
  ``The Large N limit of superconformal field theories and supergravity,''
  Adv.\ Theor.\ Math.\ Phys.\  {\bf 2}, 231 (1998)
  [hep-th/9711200].

\bibitem{Gubser:1998bc} 
  S.~S.~Gubser, I.~R.~Klebanov and A.~M.~Polyakov,
  ``Gauge theory correlators from noncritical string theory,''
  Phys.\ Lett.\ B {\bf 428}, 105 (1998)
  [hep-th/9802109].

\bibitem{Witten:1998qj} 
  E.~Witten,
  ``Anti-de Sitter space and holography,''
  Adv.\ Theor.\ Math.\ Phys.\  {\bf 2}, 253 (1998)
  [hep-th/9802150].

\bibitem{Aldazabal:2000sa}
  G.~Aldazabal, L.~E.~Ibanez, F.~Quevedo, A.~M.~Uranga,
  ``D-branes at singularities: A Bottom up approach to the string embedding of the standard model,''
  JHEP {\bf 0008} (2000) 002
  [hep-th/0005067].

\bibitem{Berenstein:2001nk} 
  D.~Berenstein, V.~Jejjala and R.~G.~Leigh,
  ``The Standard model on a D-brane,''
  Phys.\ Rev.\ Lett.\  {\bf 88}, 071602 (2002)
  [hep-ph/0105042].

\bibitem{Verlinde:2005jr} 
  H.~Verlinde and M.~Wijnholt,
  ``Building the standard model on a D3-brane,''
  JHEP {\bf 0701}, 106 (2007)
  [hep-th/0508089].

\bibitem{Hanany:2005ve} 
  A.~Hanany and K.~D.~Kennaway,
  ``Dimer models and toric diagrams,''
  hep-th/0503149.

\bibitem{Franco:2005rj} 
  S.~Franco, A.~Hanany, K.~D.~Kennaway, D.~Vegh and B.~Wecht,
  ``Brane dimers and quiver gauge theories,''  JHEP {\bf 0601}, 096 (2006)  [hep-th/0504110].  

\bibitem{Franco:2005sm} 
  S.~Franco, A.~Hanany, D.~Martelli, J.~Sparks, D.~Vegh and B.~Wecht,
  ``Gauge theories from toric geometry and brane tilings,''
  JHEP {\bf 0601}, 128 (2006)
  [hep-th/0505211].

\bibitem{Franco:2006es} 
  S.~Franco and A.~M .Uranga,
  ``Dynamical SUSY breaking at meta-stable minima from D-branes at obstructed geometries,''  JHEP {\bf 0606}, 031 (2006)  [hep-th/0604136].  

\bibitem{Forcella:2008au} 
  D.~Forcella, I.~Garcia-Etxebarria and A.~Uranga,
  ``E3-brane instantons and baryonic operators for D3-branes on toric singularities,''  JHEP {\bf 0903}, 041 (2009)  [arXiv:0806.2291 [hep-th]].  

\bibitem{Franco:2012mm} 
  S.~Franco,
  ``Bipartite Field Theories: from D-Brane Probes to Scattering Amplitudes,''  arXiv:1207.0807 [hep-th].  

\bibitem{Xie:2012mr} 
  D.~Xie and M.~Yamazaki,
  ``Network and Seiberg Duality,''  JHEP {\bf 1209}, 036 (2012)  [arXiv:1207.0811 [hep-th]].  

\bibitem{Franco:2013pg} 
  S.~Franco,
  ``Cluster Transformations from Bipartite Field Theories,''
  arXiv:1301.0316 [hep-th].

\bibitem{GK}
A.~Goncharov and R.~Kenyon, 
``Dimers and cluster integrable systems,''
arXiv:1107.5588 [math.AG]

\bibitem{Franco:2011sz} 
  S.~Franco,
  ``Dimer Models, Integrable Systems and Quantum Teichmuller Space,''
  JHEP {\bf 1109}, 057 (2011)
  [arXiv:1105.1777 [hep-th]].

\bibitem{Eager:2011dp} 
  R.~Eager, S.~Franco and K.~Schaeffer,
  ``Dimer Models and Integrable Systems,''
  JHEP {\bf 1206}, 106 (2012)
  [arXiv:1107.1244 [hep-th]].

\bibitem{Franco:2012hv} 
  S.~Franco, D.~Galloni and Y.~-H.~He,
  ``Towards the Continuous Limit of Cluster Integrable Systems,''
  JHEP {\bf 1209}, 020 (2012)
  [arXiv:1203.6067 [hep-th]].

\bibitem{Amariti:2012dd} 
  A.~Amariti, D.~Forcella and A.~Mariotti,
  ``Integrability on the Master Space,''
  JHEP {\bf 1206}, 053 (2012)
  [arXiv:1203.1616 [hep-th]].

\bibitem{ArkaniHamed:2012nw} 
  N.~Arkani-Hamed, J.~L.~Bourjaily, F.~Cachazo, A.~B.~Goncharov, A.~Postnikov and J.~Trnka,
  ``Scattering Amplitudes and the Positive Grassmannian,''
  arXiv:1212.5605 [hep-th].

\bibitem{Franco:2012wv} 
  S.~Franco, D.~Galloni and R.~-K.~Seong,
  ``New Directions in Bipartite Field Theories,''
  arXiv:1211.5139 [hep-th].

\bibitem{Amariti:2013ija} 
  A.~Amariti and D.~Forcella,
  ``Scattering Amplitudes and Toric Geometry,''
  arXiv:1305.5252 [hep-th].

\bibitem{Heckman:2012jh} 
  J.~J.~Heckman, C.~Vafa, D.~Xie and M.~Yamazaki,
  ``String Theory Origin of Bipartite SCFTs,''
  JHEP {\bf 1305}, 148 (2013)
  [arXiv:1211.4587 [hep-th]].

\bibitem{Feng:2005gw} 
  B.~Feng, Y.~-H.~He, K.~D.~Kennaway and C.~Vafa,
  ``Dimer models from mirror symmetry and quivering amoebae,''  Adv.\ Theor.\ Math.\ Phys.\  {\bf 12}, 489 (2008)  [hep-th/0511287].  

\bibitem{Aharony:1997ju} 
  O.~Aharony, A.~Hanany and ,
  ``Branes, superpotentials and superconformal fixed points,''  Nucl.\ Phys.\ B {\bf 504}, 239 (1997)  [hep-th/9704170].  

\bibitem{Aharony:1997bh} 
  O.~Aharony, A.~Hanany, B.~Kol and ,
  ``Webs of (p,q) five-branes, five-dimensional field theories and grid diagrams,''  JHEP {\bf 9801}, 002 (1998)  [hep-th/9710116].  

\bibitem{Krippendorf:2010hj}
  S.~Krippendorf, M.~J.~Dolan, A.~Maharana and F.~Quevedo,
  ``D-branes at Toric Singularities: Model Building, Yukawa Couplings and Flavour Physics,''
ÊÊJHEP {\bf 1006} (2010) 092
ÊÊ[arXiv:1002.1790 [hep-th]].
ÊÊ

\bibitem{Dolan:2011qu}
  M.~J.~Dolan, S.~Krippendorf and F.~Quevedo,
  ``Towards a Systematic Construction of Realistic D-brane Models on a del Pezzo Singularity,''
ÊÊJHEP {\bf 1110} (2011) 024
ÊÊ[arXiv:1106.6039 [hep-th]].
ÊÊ

\bibitem{Cicoli:2013zha}
  M.~Cicoli, S.~Krippendorf, C.~Mayrhofer, F.~Quevedo and R.~Valandro,
  ``The Web of D-branes at Singularities in Compact Calabi-Yau Manifolds,''
ÊÊJHEP {\bf 1305} (2013) 114
ÊÊ[arXiv:1304.2771 [hep-th]].
ÊÊ

\bibitem{Ouyang:2003df} 
  P.~Ouyang,
  ``Holomorphic D7 branes and flavored N=1 gauge theories,''
  Nucl.\ Phys.\ B {\bf 699}, 207 (2004)
  [hep-th/0311084].

\bibitem{Kuperstein:2004hy}
  S.~Kuperstein,
  ``Meson spectroscopy from holomorphic probes on the warped deformed conifold,''
ÊÊJHEP {\bf 0503} (2005) 014
ÊÊ[hep-th/0411097].
ÊÊ

\bibitem{Karch:2002sh}
  A.~Karch and E.~Katz,
  ``Adding flavor to AdS / CFT,''
  JHEP {\bf 0206} (2002) 043
  [hep-th/0205236].

\bibitem{Klebanov:1998hh}
  I.~R.~Klebanov and E.~Witten,
  ``Superconformal field theory on three-branes at a Calabi-Yau singularity,''
  Nucl.\ Phys.\ B {\bf 536} (1998) 199
  [hep-th/9807080].

\bibitem{Morrison:1998cs}
  D.~R.~Morrison and M.~R.~Plesser,
  ``Nonspherical horizons. 1.,''
  Adv.\ Theor.\ Math.\ Phys.\  {\bf 3} (1999) 1
  [hep-th/9810201].

\bibitem{Gubser:1998fp} 
  S.~S.~Gubser and I.~R.~Klebanov,
  ``Baryons and domain walls in an N=1 superconformal gauge theory,''
  Phys.\ Rev.\ D {\bf 58}, 125025 (1998)
  [hep-th/9808075].

\bibitem{Berenstein:2002ke} 
  D.~Berenstein, C.~P.~Herzog and I.~R.~Klebanov,
  ``Baryon spectra and AdS /CFT correspondence,''
  JHEP {\bf 0206}, 047 (2002)
  [hep-th/0202150].

\bibitem{Beasley:2002xv}
  C.~E.~Beasley,
  ``BPS branes from baryons,''
  JHEP {\bf 0211} (2002) 015
  [hep-th/0207125].

\bibitem{Butti:2006au}
  A.~Butti, D.~Forcella and A.~Zaffaroni,
  ``Counting BPS baryonic operators in CFTs with Sasaki-Einstein duals,''
  JHEP {\bf 0706} (2007) 069
  [hep-th/0611229].
  
\bibitem{Forcella:2007wk}
  D.~Forcella, A.~Hanany and A.~Zaffaroni,
  ``Baryonic Generating Functions,''
  JHEP {\bf 0712} (2007) 022
  [hep-th/0701236 [HEP-TH]].

\bibitem{Leigh:1998hj}
  R.~G.~Leigh, M.~Rozali,
  ``Brane boxes, anomalies, bending and tadpoles,''
  Phys.\ Rev.\ D {\bf 59} (1999) 026004
  [hep-th/9807082].

\bibitem{Wijnholt:2002qz}
  M.~Wijnholt,
  ``Large volume perspective on branes at singularities,''
  Adv.\ Theor.\ Math.\ Phys.\  {\bf 7} (2004) 1117
  [hep-th/0212021].
  
\bibitem{Herzog:2003zc}
  C.~P.~Herzog,
  ``Exceptional collections and del Pezzo gauge theories,''
  JHEP {\bf 0404} (2004) 069
  [hep-th/0310262].
  
\bibitem{Hanany:2006nm}
  A.~Hanany, C.~P.~Herzog and D.~Vegh,
  ``Brane tilings and exceptional collections,''
  JHEP {\bf 0607} (2006) 001
  [hep-th/0602041].

\bibitem{Uranga:2002pg}
  A.~M.~Uranga,
  ``Local models for intersecting brane worlds,''
  JHEP {\bf 0212} (2002) 058
  [hep-th/0208014].

\bibitem{Uranga:2000xp}
  A.~M.~Uranga,
  ``D-brane probes, RR tadpole cancellation and K theory charge,''
  Nucl.\ Phys.\ B {\bf 598} (2001) 225
  [hep-th/0011048].

\bibitem{Ibanez:1998qp}
  L.~E.~Ibanez, R.~Rabadan, A.~M.~Uranga and ,
  ``Anomalous U(1)'s in type I and type IIB D = 4, N=1 string vacua,''
  Nucl.\ Phys.\ B {\bf 542} (1999) 112
  [hep-th/9808139].

\bibitem{Aldazabal:1999nu}
  G.~Aldazabal, D.~Badagnani, L.~E.~Ibanez, A.~M.~Uranga and ,
  ``Tadpole versus anomaly cancellation in D = 4, D = 6 compact IIB orientifolds,''
  JHEP {\bf 9906} (1999) 031
  [hep-th/9904071].

\bibitem{Cremonesi:2013aba} 
  S.~Cremonesi, A.~Hanany and R.~-K.~Seong,
  ``Double Handled Brane Tilings,''
  arXiv:1305.3607 [hep-th].

\bibitem{Park:1999ep}
  J.~Park, R.~Rabadan and A.~M.~Uranga,
  ``Orientifolding the conifold,''
  Nucl.\ Phys.\ B {\bf 570} (2000) 38
  [hep-th/9907086].
  
\bibitem{Uranga:1998vf}
  A.~M.~Uranga,
  ``Brane configurations for branes at conifolds,''
  JHEP {\bf 9901} (1999) 022
  [hep-th/9811004].

\bibitem{Dasgupta:1998su}
  K.~Dasgupta and S.~Mukhi,
  ``Brane constructions, conifolds and M theory,''
  Nucl.\ Phys.\ B {\bf 551} (1999) 204
  [hep-th/9811139].


\bibitem{Brodie:1997sz}
  J.~H.~Brodie and A.~Hanany,
  ``Type IIA superstrings, chiral symmetry, and N=1 4-D gauge theory dualities,''
  Nucl.\ Phys.\ B {\bf 506} (1997) 157
  [hep-th/9704043].

\bibitem{Hanany:1997sa}
  A.~Hanany and A.~Zaffaroni,
  ``Chiral symmetry from type IIA branes,''
  Nucl.\ Phys.\ B {\bf 509} (1998) 145
  [hep-th/9706047].
    
\bibitem{Ganor:1996pe}
  O.~J.~Ganor,
  ``A Note on zeros of superpotentials in F theory,''
  Nucl.\ Phys.\ B {\bf 499} (1997) 55
  [hep-th/9612077].

\bibitem{Buican:2008qe}
  M.~Buican and S.~Franco,
  ``SUSY breaking mediation by D-brane instantons,''
ÊÊJHEP {\bf 0812} (2008) 030
ÊÊ[arXiv:0806.1964 [hep-th]].
ÊÊ

\bibitem{Baumann:2006th}
  D.~Baumann, A.~Dymarsky, I.~R.~Klebanov, J.~M.~Maldacena, L.~P.~McAllister and A.~Murugan,
  ``On D3-brane Potentials in Compactifications with Fluxes and Wrapped D-branes,''
  JHEP {\bf 0611} (2006) 031
  [hep-th/0607050].

\bibitem{Baumann:2007np}
  D.~Baumann, A.~Dymarsky, I.~R.~Klebanov, L.~McAllister and P.~J.~Steinhardt,
  ``A Delicate universe,''
  Phys.\ Rev.\ Lett.\  {\bf 99} (2007) 141601
  [arXiv:0705.3837 [hep-th]].

\bibitem{Marchesano:2009rz}
  F.~Marchesano and L.~Martucci,
  ``Non-perturbative effects on seven-brane Yukawa couplings,''
  Phys.\ Rev.\ Lett.\  {\bf 104} (2010) 231601
  [arXiv:0910.5496 [hep-th]].

\bibitem{Font:2012wq}
  A.~Font, L.~E.~Ibanez, F.~Marchesano and D.~Regalado,
  ``Non-perturbative effects and Yukawa hierarchies in F-theory SU(5) Unification,''
  JHEP {\bf 1303} (2013) 140
  [arXiv:1211.6529 [hep-th]].
  
\bibitem{ArkaniHamed:2001ie} 
  N.~Arkani-Hamed, A.~G.~Cohen, D.~B.~Kaplan, A.~Karch and L.~Motl,
  ``Deconstructing (2,0) and little string theories,''
  JHEP {\bf 0301}, 083 (2003)
  [hep-th/0110146].

\bibitem{Bah:2011vv} 
  I.~Bah, C.~Beem, N.~Bobev and B.~Wecht,
  ``AdS/CFT Dual Pairs from M5-Branes on Riemann Surfaces,''
  Phys.\ Rev.\ D {\bf 85}, 121901 (2012)
  [arXiv:1112.5487 [hep-th]].

\bibitem{Bah:2012dg} 
  I.~Bah, C.~Beem, N.~Bobev and B.~Wecht,
  ``Four-Dimensional SCFTs from M5-Branes,''
  JHEP {\bf 1206}, 005 (2012)
  [arXiv:1203.0303 [hep-th]].

  \end{thebibliography}
\end{document}